\documentclass[prd,twocolumn,reprint,preprintnumbers,nofootinbib,superscriptaddress]{revtex4-2}
\input{header.tex}

\definecolor{color_git}{rgb}{0.098, 0.160, 0.345}
\newcommand{\gitlink}{\href{https://github.com/MiguelEA/BBNEasyALP}{\textsc{g}it\textsc{h}ub {\large\color{color_git}\faGithub}}$\,$}

\begin{document}

\reportnum{-2}{CERN-TH-2025-222}
\title{Nucleosynthesis and CMB bounds on photophilic ALPs: a fresh look}

\author{Miguel Escudero Abenza}
\email{miguel.escudero@cern.ch}
\affiliation{Theoretical Physics Department, CERN, 1211 Geneva 23, Switzerland}
\author{Clara Garcia-Perez}
\email{claragarperez@gmail.com}
\affiliation{Facultad de Ciencias, Universidad de Zaragoza, E-50009 Zaragoza, Spain}
\affiliation{Theoretical Physics Department, CERN, 1211 Geneva 23, Switzerland}
\author{Maksym Ovchynnikov}
\email{maksym.ovchynnikov@cern.ch}
\affiliation{Theoretical Physics Department, CERN, 1211 Geneva 23, Switzerland}

\date{\today}

\begin{abstract}
We provide a fresh look at the cosmological constraints on axion-like particles (ALPs) that couple predominantly to photons, focusing on lifetimes $\tau_{a} \lesssim 10^{4}\, {\rm s}$ and masses $m_a\lesssim 10\,{\rm GeV}$. We consider Big Bang Nucleosynthesis (BBN) and Cosmic Microwave Background (CMB) bounds and explore how these limits depend upon the unknown reheating temperature of the Universe, $T_{\rm reh}$. Compared with some previous studies, we account for the rare decays of these ALPs into light hadrons and show that this leads to extended constraints for several reheating temperatures. Our limits are cast in a model-independent way, and we identify regions of parameter space where these ALPs could alleviate small tensions in the determinations of $N_{\rm eff}$ and the deuterium abundance. Our Mathematica BBN code \texttt{BBNEasyALP} is publicly available at~\gitlink.
\end{abstract}

\maketitle

\section{Introduction}
\label{sec:introduction}

The QCD axion is an inevitable consequence of the Peccei-Quinn solution~\cite{Peccei:1977ur,Peccei:1977hh} to the strong CP problem~\cite{Weinberg:1977ma,Wilczek:1977pj}. Their broader counterparts, axion-like particles (ALPs), constitute a natural generalization of this idea and are generically expected to appear in string theory~\cite{Arvanitaki:2009fg,Demirtas:2021gsq}. {They are pseudoscalars coupled to Standard Model fields bilinears:
\begin{equation}
    \mathcal{L}_{a} = \frac{g_{aGG}}{4}a\, G_{\mu\nu}\tilde{G}^{\mu\nu}+ \frac{g_{a\gamma\gamma}}{4}a\, F_{\mu\nu}F^{\mu\nu} + \dots,
\end{equation}
where $a$ is the ALP field, and $g_{aGG/\gamma\gamma}$ are dimensionful couplings.
}

Although the landscape of string theory vacua is exceedingly rich~\cite{McAllister:2023vgy}, there is an active effort to derive concrete predictions for the masses and couplings of ALPs in controlled corners of various string theory frameworks~\cite{Cicoli:2012sz,Gendler:2023kjt,Sheridan:2024vtt,Leedom:2025mlr,Reig:2025dqb,Benabou:2025kgx}. This theoretical motivation is further reinforced by the fact that ALPs over a broad range of masses and interaction strengths can be probed by laboratory searches, astrophysical observations, and cosmological measurements; see, e.g.,~\cite{Jaeckel:2010ni,Beacham:2019nyx,Irastorza:2018dyq,Sikivie:2020zpn,Caputo:2024oqc,deBlas:2025gyz} for reviews.

In this work, we concentrate on the cosmological constraints that can be derived on ALPs that couple predominantly to photons. We are motivated by three main reasons: 1) the recent interest in ALPs arising from low-energy string theory realizations that feature ALPs with very different masses, couplings, and primordial abundances, 2) the fact that cosmological constraints on these ALPs are among the strongest across wide regions of parameter space, and 3) that the cosmological analysis of photophilic ALPs can still be refined.

In particular, these cosmological bounds have been studied in detail by several groups in the past~\cite{Masso:1995tw,Masso:1997ru,Cadamuro:2010cz,Cadamuro:2011fd,Cadamuro:2012rm,Millea:2015qra,Depta:2020wmr,Langhoff:2022bij,Yin:2025amn}, with the results from Refs.~\cite{Cadamuro:2011fd} and~\cite{Depta:2020wmr} adopted by the Particle Data Group~\cite{ParticleDataGroup:2024cfk}. While these studies comprehensively scanned the ALP parameter space, none of them \emph{simultaneously} precisely calculated \neff, included all effects from rare ALP decays into mesons, and explored the cosmological model dependency of the bounds. This is precisely the gap we fill in this work: we take these physical effects into account and present results for various reheating temperatures, ranging from $\Treh \simeq  5\mev$ to $\Treh \simeq 10^{15}\gev$.

We revisit cosmological constraints on MeV–GeV axion-like particles (ALPs) that couple predominantly to photons and have lifetimes $\tau_a \lesssim 10^4\,\mathrm{s}$. We focus on this regime because, for longer lifetimes, the bounds are dominated by photodisintegration of light nuclei, which has been treated accurately in Ref.~\cite{Depta:2020wmr}, see~\cite{Kawasaki:1994sc,Cyburt:2002uv,Kawasaki:2004qu,Kawasaki:2017bqm} for previous studies. Relative to Ref.~\cite{Cadamuro:2011fd}, we obtain substantially stronger and broader exclusions, driven primarily by improved cosmological data and refined modeling. Compared with Ref.~\cite{Depta:2020wmr}, our limits are order-of-magnitude consistent at very high reheating temperatures, but we identify new excluded regions for $m_a \gtrsim 400\mev$ even for moderate reheating temperatures, $\Treh\sim 1\,\mathrm{TeV}$. The origin is that, although ALPs decay mainly into photons, once $m_a>2m_{\pi^{\pm}}$, a subdominant hadronic channel into light mesons opens; these mesons undergo strong interactions in the plasma, driving the neutron-to-proton ratio above its Standard Big-Bang Nucleosynthesis (SBBN) value and thereby increasing the primordial helium yield.

\begin{figure*}[t!]
    \centering
    \includegraphics[width=0.496\linewidth]{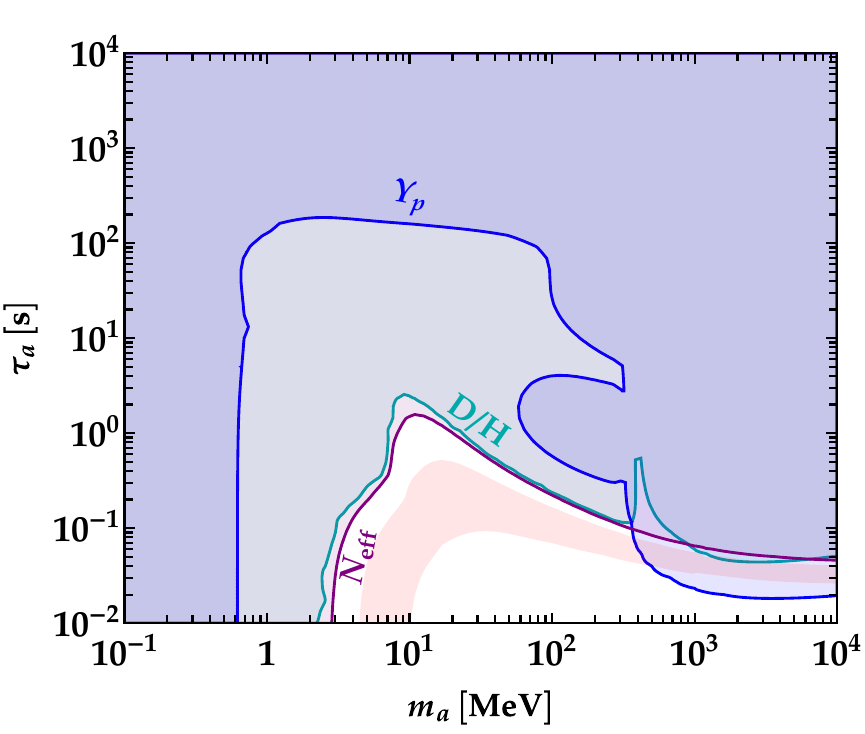}~\includegraphics[width=0.504\linewidth]{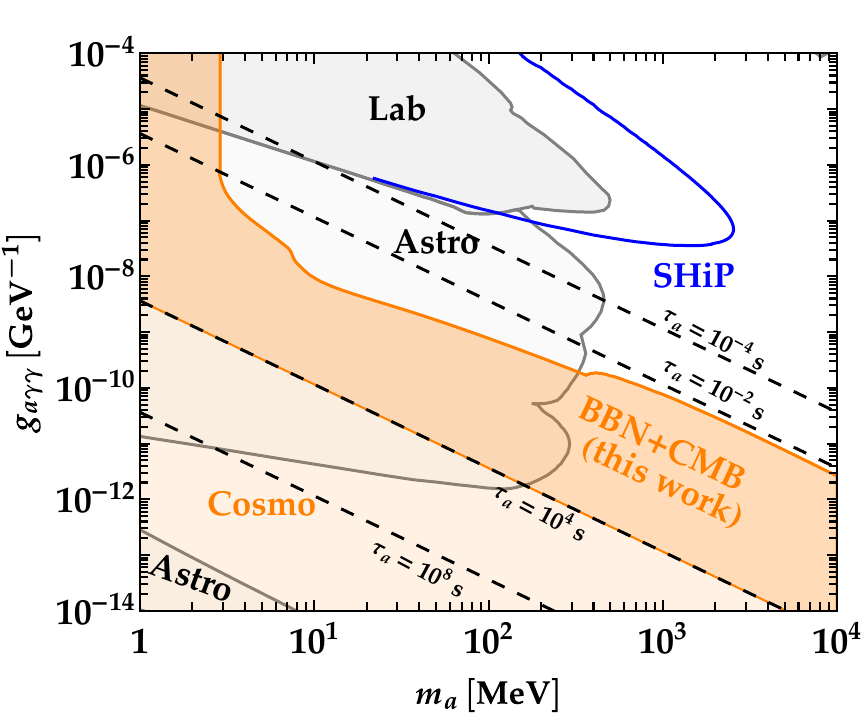}
    \caption{ALP parameter space. \textit{Left panel}: plane mass-lifetime, showing cosmological constraints in the scenario with a high reheating temperature $T_{\rm reh}\ge 10^{10}\,{\rm GeV}$ (see Fig.~\ref{fig:results-ALPs-reheating-temperature} for other choices of $T_{\rm reh}$). The blue domain shows the constraints from primordial helium-4 abundance observations ($Y_{\rm P}$), the cyan one -- from the primordial deuterium abundance (${\rm D/H}|_{\rm P}$), while the purple one corresponds to the bounds from \neff measurements. The light-red band corresponds to the range of masses and lifetimes where ALPs cause $\neff = 2.81\pm 0.12$, preferred by the latest CMB measurements (see Sec.~\ref{sec:methodology} for details).  \textit{Right panel}: the results in the plane of ALP mass and coupling $g_{a\gamma\gamma}${, where the latter is related to the lifetime via $\tau_a \propto g_{a\gamma\gamma}^{-2}$ (see Eq.~\eqref{eq:ALP-lifetime}). For convenience, by dashed black lines, we show the iso-contours of fixed ALP lifetime.} In addition to the cosmological bounds, we also depict astrophysical constraints (see~\cite{Jaeckel:2017tud,Caputo:2022mah,Diamond:2023scc,Diamond:2023cto,Fiorillo:2025yzf,Candon:2025ypl}), the domain excluded by laboratory experiments~\cite{deBlas:2025gyz}, as well as the sensitivity of the recently approved SHiP experiment~\cite{SHiP:2025ows}, whose projected reach is among the leading probes of long-lived, GeV-mass ALPs~\cite{deBlas:2025gyz}. The results for $\tau_{a}<10^{4}\s$ are as obtained in this work. The larger lifetimes are excluded by the combination of various bounds emerging on electromagnetically decaying thermal relics~\cite{Poulin:2016anj,Kawasaki:2017bqm,Langhoff:2022bij}. {The upper boundary of the cosmological exclusion is set by whether ALPs can be present in the plasma at temperatures $T\lesssim{\rm few~MeV}$. For masses $m_a \gtrsim 1~{\rm MeV}$, this requires sufficiently late decays, $\tau_a \gtrsim 0.02~{\rm s}$ (and hence smaller couplings), so that relic ALPs survive until this epoch. Lighter ALPs can be abundantly produced in the MeV plasma, so their presence at $T\sim{\rm MeV}$ does not require $\tau_a \gtrsim 0.02~{\rm s}$; therefore, the constraints extend to much larger couplings.}}
    \label{fig:ALP-results}
\end{figure*}

The core results of our analysis are shown in Figs.~\ref{fig:ALP-results}-\ref{fig:results-Y-tau-plane}, and will be elaborated upon in detail throughout our study.\footnote{The approach we developed may be applied to generic particles decaying in the MeV plasma and later. {A version of our Mathematica BBN code \texttt{BBNEasyALP} is publicly available at~\gitlink.}} The panel on the left shows the BBN bounds from helium and deuterium, and $\neff$ constraints from the CMB assuming a very high reheating temperature $\Treh \simeq 10^{10}\gev$ in the plane of ALP lifetime versus mass. The right panel shows the results in the plane of the axion-photon coupling and mass in the context of an array of laboratory and astrophysical constraints. Importantly, we will also report our results in a model-independent fashion so that they can easily be recast for other photophilic/electrophilic light relics. 

The remainder of our study is structured as follows. First, in Section~\ref{sec:cosmoeffects}, we provide an overview of the main cosmological effects on decaying ALPs during BBN and for the CMB. Then, in Section~\ref{sec:methodology}, we outline our calculation of the early Universe thermodynamics in the presence of ALPs, including their production, decay, and impact on the primordial element abundances and the CMB spectra. In Section~\ref{sec:results}, we present the resulting bounds on the parameter space and compare with previous studies. We conclude in Section~\ref{sec:conclusions}. The interested reader is referred to Appendices~\ref{app:alp-production}-\ref{app:BBN-chain} discussing in-depth our methodology, and~\ref{app:some-results},~\ref{app:previous-works-comparison} devoted to comparison with previous works.

\section{ALPs coupled to electromagnetism and their cosmological implications}
\label{sec:cosmoeffects}

We consider an axion-like particle coupled to electromagnetism via the usual $F\tilde{F}$ operator that typically arises from interactions of this ALP with heavy fermions featuring anomalous charges. Its interaction Lagrangian at low energies $E\ll M_W$ is:
\begin{align}
\label{eq:Laint}
    \mathcal{L}_{a} = \frac{g_{a\gamma\gamma}}{4}a F_{\mu\nu}\tilde{F}^{\mu\nu}\,.
\end{align}
with $a$ being the ALP field, $F,\tilde{F}$ the photon strength and dual strength, corresponding, and $g_{a\gamma\gamma}$ the coupling constant. 

We will consider that this is the dominant coupling for our ALP and as such, the $a\to \gamma \gamma$ decay controls the ALP lifetime:
\begin{align}
    \tau_a &= \left(\frac{g_{a\gamma \gamma}^2 m_a^3}{64\pi}\right)^{-1}\,, \label{eq:ALP-lifetime}\\
           &\simeq 130\,{\rm s}\, \left(\frac{10^{-9}\gev^{-1}}{g_{a\gamma\gamma}}\right)^{2} \left(\frac{10\mev}{m_a}\right)^3\,.\nonumber
\end{align}
The $g_{a\gamma\gamma}$ coupling implies a cosmological ALP production primarily via two processes: $f^\pm \gamma \to f^\pm a$, where $f$ is any charged fermion and $\gamma \gamma \to a$, both known as Primakoff and coalescence processes, respectively. In particular, the rate for the first one roughly scales as $\Gamma \simeq 10^{-3} \, g_{a\gamma \gamma}^2 T^3 $ quite independently of the ALP mass, and it typically implies that ALPs will be in thermal equilibrium in the early Universe down to temperatures
\begin{align}\label{eq:ALP_Tdecoupling}
    T_{\rm a}^{\rm dec}\simeq 200\gev \,\left(\frac{m_a}{10\mev}\right)^3 \,\left(\frac{\tau_a}{100\,{\rm s}}\right) \sqrt{\frac{g_\star}{100}}\,\,,
\end{align}
where $g_{\star}$ is the number of effective relativistic species at the time. 

From this expression, one can easily then see that these species will be produced with (large) thermal abundances during the hot thermal stage after the Big Bang. Critically, this means that they will still have large densities at $T\simeq 1\mev$ when the SM weak interactions stop being efficient, the neutron abundance in the Universe is set, and the cosmic neutrino background forms. 

The cosmological implications of ALPs depend upon their mass and lifetime, but can roughly be divided into two distinct mechanisms. First, they will, in general, modify the expansion history of the Universe from their contribution to its energy density. If this happens after the decoupling of the weak interactions at $T\lesssim 2\mev$, it will lead to an impact on the number of effective relativistic neutrino species ($\neff$) and to the synthesis of the primordial elements during BBN. Second, the ALP's decay products may directly alter the light nuclei' abundances as synthesized during BBN.

Critically, the second consequence depends strongly upon the final decay products of the ALP. In our case, while by definition we have considered that our ALPs interact exclusively with two photons,\footnote{Additional interactions, similar to those of the hadronically coupled ALPs, may appear because of the renormalization group flow running from the scale $\Lambda$, at which the Lagrangian~\eqref{eq:Laint} is defined, to the scales $Q =  m_{a}$ at which the decay rates are defined~\cite{Bauer:2020jbp}. However, using the framework from Ref.~\cite{Ovchynnikov:2025gpx}, we have found that these $\Lambda$-dependent interactions do not significantly change our results; in particular, the RG-flow-induced hadronic decay modes have branching ratios of the order of those mediated at tree-level by the photonic coupling $g_{a\gamma\gamma}$. Therefore, we conservatively do not include them.} it does not imply that $a\to \gamma \gamma$ is their only possible decay mode. For large $m_a$, the ALPs will decay via off-shell photons into $\gamma e^+e^-$, $\gamma \mu^+\mu^-$, and $\gamma \pi^+\pi^-$, among other final states, depending upon $m_a$. All these channels would have small branching ratios, typically at the ${\rm Br}\sim 10^{-3}$ level. Importantly, for the case of mesons in the final state, since these particles interact strongly and are relatively long-lived ($\tau_{\pi^\pm} \simeq 10^{-8}\,{\rm s}$), they can lead to key modifications of the nuclear reaction network compared with the Standard Model case. In what follows, we discuss in detail these two cosmological implications.

\subsection{Modification of the Universe's thermal history}

We have robust indirect evidence for the particle content of the Universe at temperatures $T \lesssim 5\mev$ and a corresponding cosmic age $t_{\rm U} \gtrsim 0.03\,{\rm s}$. Within the Standard Cosmological Scenario, in this epoch, the plasma consisted of thermal $e^\pm$, photons, and neutrinos with equilibrium number densities, $n_i \propto T^3$. In addition, a small but cosmologically crucial baryon asymmetry was present, quantified today by the baryon-to-photon ratio $\eta_B \equiv n_b/n_\gamma = (6.14 \pm 0.04) \times 10^{-10}$~\cite{Planck:2018vyg}.
 
As the Universe cools, various key physical processes take place: (i) at $T\simeq 2\mev$, neutrinos stop interacting with the rest of the plasma and from then onwards they simply free-stream; (ii) at $T_\gamma\simeq 0.7\mev$, the weak interactions interconverting protons and neutrons in the early Universe freeze out, setting a primordial neutron-to-proton density ratio of $\sim 1/6$; (iii) at $T\lesssim m_e$, electrons and positrons annihilate, heating up the photons relative to neutrinos and yielding $T_\gamma/T_\nu \simeq 1.4$ and $\neff\simeq 3.04$; (iv) at $T_\gamma \simeq 0.075\mev$, deuterium becomes stable against photodissociation, and quite rapidly almost all the neutrons in the Universe form ${}^4 {\rm He}$, and the process leads to residual fractions of deuterium and ${}^3\mathrm{He}$ at the $\sim 10^{-5}$ level and through the process a fraction of $\sim 10^{-10}$ of ${}^7\mathrm{Li}$ is synthesized.

As we will discuss in Sec.~\ref{sec:methodology}, this picture is corroborated by precision measurements of $\neff$ and the abundances of helium-4 and deuterium. But how do ALPs alter it? 

First, the presence of an ALP would lead to a modification of the time-temperature relation as ALPs directly contribute to the expansion rate of the Universe, $H = \sqrt{8\pi \rho_{\rm tot}/(3m_{\rm pl}^2)}$ where $\rho_{\rm tot}$ is the energy density of the Universe and $m_{\rm pl} = 1.22\times 10^{19}\,{\rm GeV}$. Second, since the ALPs we consider decay into photons, they will alter the ratio $T_\gamma/T_\nu$, and this in turn leads to a lower value of $\neff$ as relevant for CMB observations. Third, even if the ALP decay takes place after BBN ($t_{\rm U} \simeq 3\,{\rm min}$) an injection of a significant number of photons will dilute the net baryon-to-photon ratio as compared with what is inferred from CMB observations and this will lead to an impact on the inferred primordial element abundances of all nuclei, but in particular for deuterium.

In section~\ref{sec:methodology}, we describe how we account for the expansion of the Universe in the presence of an ALP and take into account all its relevant interactions and effects. 

\subsection{Modification of the BBN reaction chain}

In the standard cosmological model, protons and neutrons interact via the weak interactions: $n + \nu_e \leftrightarrow p + e^- $, $n + e^+\leftrightarrow p + \bar{\nu}_e $, $n \leftrightarrow p + e^-+ \bar{\nu}_e$. ALPs can modify the rates for these processes because they inject photons (and hence modify the $T_\gamma/T_\nu$ temperature ratio), but also can induce new channels. 

In particular, ALPs with $m_a > 2 m_{\pi^\pm}$ can and will decay some of the time to light mesons via an off-shell photon, e.g., $a\to \gamma \gamma^\star \to \gamma \pi^+\pi^-$. Unlike electrons and neutrinos, these particles interact \textit{strongly} with nucleons and, in particular, can lead to a huge impact on the neutron-to-proton ratio in the Universe and on the abundances of various light nuclei~\cite{Reno:1987qw,Kawasaki:2000en,Kawasaki:2004qu,Pospelov:2010cw,Akita:2024ork,Omar:2025jue}. For instance, pions can convert nucleons in the following way : 
\begin{subequations}\label{eq:pn-conversion-mesons}
\begin{align}    
    \pi^{-}+p&\to \pi^{0}/\gamma + n \,,\, \dots\\
    \pi^{+}+n&\to \pi^{0}/\gamma + p \,,\, \dots
\end{align}
\end{subequations}
where $\dots$ mean processes with higher multiplicities. The characteristic cross-section for these processes is of order
$\sigma \simeq 1/m_\pi^{2}$. The net \pn exchange, however, is biased from protons to neutrons, both because protons are more abundant in the plasma and because interactions of $\pi^{-}$ with charged SM plasma particles are Coulomb-enhanced. Consequently, even though charged pions are present in the plasma only for a very short time, $\tau_{\pi^\pm} \simeq 2 \times 10^{-8}\s$, they can still
significantly modify the number densities of light nuclei: they interact at a rate
\begin{equation}
\frac{\sigma_{\pi}}{\sigma_{\nu}} \simeq \frac{1}{m_\pi^{2} G_F^{2} T^{2}}
\simeq 10^{17} \left(\frac{{\rm MeV}}{T}\right)^{2}\,,
\end{equation}
times faster than neutrinos or electrons. As we will see below, this typically implies that the hadronic branching ratio of our ALPs must be very small, ${\rm Br}(a \to {\rm hadrons}) \lesssim 10^{-5}$, or else their primordial abundance must be suppressed.

Furthermore, if the ALPs have lifetimes $\tau_{a}\gtrsim 100\s$, once significant abundances of ${}^4{\rm He}$ have formed, the mesons produced in their decays may additionally participate in the nuclear dissociation reactions of the type
\begin{equation}
\pi^{-}+ ^{4}\text{He}\to t + n, \dots
\label{eq:dissociation}
\end{equation}
This type of process will, in turn, also lead to an increase in deuterium, as the neutrons produced will be readily captured by the many protons in the plasma: $p + n \to D + \gamma$.

\section{Methodology}
\label{sec:methodology}

In this section, we present our entire pipeline: we discuss our calculation of the primordial ALP abundance given a reheating temperature, the calculation of their branching ratio into hadrons, our modeling of the expansion history of the Universe in their presence, the impact of their decays in the BBN network, and the cosmological data we use to contrast our predictions against observations. The practitioner is referred to the appendices~\ref{app:appendices}, where an array of technicalities is presented and discussed in detail. Our Mathematica BBN code \texttt{BBNEasyALP} is publicly available at~\gitlink.

\textbf{ALP production in the early Universe.} To obtain the evolution of the ALP population in the early Universe at temperatures $T\gg 1\mev$, we considered the integrated Boltzmann equation on the ALP abundance $\mathcal{Y}_{a}(T) \equiv n_{a}/s$, with $n_{a}$ being the ALP number density and $s$ the entropy density of the Universe:
\begin{equation}
\frac{d \mathcal{Y}_{a}}{dt} = \Gamma_{a}(\mathcal{Y}_{a,\text{eq}}-\mathcal{Y}_{a})\,,
\label{eq:ALP-abundance-equation}
\end{equation}
$\Gamma_{a}$ is the ALP production rate. The relevant processes of interest are the Primakoff scattering $\gamma+f^\pm\to a+f^\pm$, where $f^\pm$ is any SM charged fermion, and the photon fusion $\gamma\gamma\to a$~\cite{Cadamuro:2011fd,Cadamuro:2012rm}. To calculate $\Gamma_{a}$, we follow the approach of Ref.~\cite{Edsjo:1997bg}, see Appendix~\ref{app:alp-production}. Namely, we approximate the distributions of the Standard Model fermions with a Maxwell-Boltzmann distribution and reduce the phase space of the ALP rate integral to one dimension. In the limit $m_{f},m_{a}\ll T$, our results for $\Gamma_{a}$ are in very good agreement with the asymptotic result from Ref.~\cite{Bolz:2000fu}.

We show results for various reheating temperatures of the Universe, which effectively means integrating Eq.~\eqref{eq:ALP-abundance-equation} from $\Treh$ considering $\mathcal{Y}_a = 0$ until $T = 20\mev$ (when we will start our BBN evolution). Since the process of ALP production is UV-dominated (namely, the production rate scales as $\Gamma \simeq 10^{-3} \,g_{a\gamma \gamma}^2\, T^3 $) the number density of ALPs will effectively be thermal (modulo some small entropy dilution) provided that $\Treh > T_{a}^{\rm dec}$ as given by Eq.~\eqref{eq:ALP_Tdecoupling}.

\textbf{ALP decay rates.} The width of the dominant decay $a\to \gamma\gamma$ is $\Gamma_{a\to \gamma\gamma} = m_{a}^{3}g_{a\gamma\gamma}^{2}/(64\pi)$. To calculate the hadronic decay palette of the ALPs, we utilized an analog of the data-driven approach from Ref.~\cite{Ilten:2018crw}; details are summarized in Appendix~\ref{app:alp-decay}. In particular, we express the widths of the exclusive processes $a\to \gamma+\text{hadrons}$ via
\begin{equation}
    \Gamma_{a\to \gamma + \text{hadrons}} \approx \int d\bar{s} \ f(\bar{s})R(\bar{s})\, ,
\end{equation}
where $\bar{s}=(p_{a}-p_{\gamma})^{2}$ is the squared invariant mass transferred to the hadrons, $f(\bar{s})$ is the ``splitting function'', while $R(\bar{s}) \equiv \sigma_{e^{+}e^{-}\to\text{hadrons}}/\sigma_{e^{+}e^{-}\to \mu^{+}\mu^{-}}$ is the experimentally measured $R$-ratio~\cite{ParticleDataGroup:2024cfk}. The resulting hadronic branching ratios are explicitly depicted in Fig.~\ref{fig:meson-multiplicities} and are $>10^{-3}$ for $m_a > 1\gev$; as we will see, even tiny values have key cosmological implications. 

\textbf{Thermodynamics of the Universe.} Decaying at MeV temperatures and later, the ALPs influence both the expansion of the Universe and BBN. However, thanks to the fact that nucleons contribute negligibly to the energy density of the Universe, these two effects can be factorized. Namely, we first solve the equations governing the expansion of the Universe in the presence of ALPs, and then use the resulting output to study the synthesis of the light elements. We provide all relevant details in Appendix~\ref{app:thermodynamics}, and, in what follows, we highlight the essential ingredients of our approach.

To study the early Universe thermodynamics, we use the integrated Boltzmann approach to solve the coupled evolution of neutrinos and the electromagnetic (EM) plasma~\cite{Escudero:2018mvt,EscuderoAbenza:2020cmq,Escudero:2025mvt}. The idea is to approximate the neutrino distribution function $f_{\nu}$ with a Fermi-Dirac distribution described by a dynamical temperature $T_{\nu}$, and then to solve the coupled system of equations for $T_{\nu}\,,T_{\gamma}\,, a$, where $T_{\gamma}$ is the temperature of the EM plasma and $a$ is the scale factor of the Universe. The approach works very well as long as neutrino spectral distortions can be neglected, which is the case here.\footnote{It has been shown that even purely electromagnetic decays may induce neutrino spectral distortions, which then shift \neff to smaller values~\cite{Akita:2024ork,Ovchynnikov:2024rfu}. However, the error $|\delta (\dneff)|$ from neglecting the distortions is typically significantly smaller than \dneff itself, and therefore this approximation is well justified.}

The important point here is that the ALPs may still be in partial equilibrium at MeV temperatures. This would happen for the ALP mass $m_{a}\lesssim 10\mev$, as we will see in Sec.~\ref{sec:results}. As a result, we need to know the ALP energy distribution throughout the evolution of the Universe. For this purpose, we use the unintegrated (Liouville) equation for the ALP evolution as in Refs.~\cite{Cadamuro:2011fd,Cadamuro:2012rm}, which we solve efficiently using the Gauss-Laguerre quadrature method for several momentum bins. We do this simultaneously with solving for $T_{\nu}\,,T_{\gamma}\,, a$, and then we have access to all relevant thermodynamic quantities. 

\textbf{Big Bang Nucleosynthesis chain.} To calculate the impact of the ALPs on BBN, we have developed a custom BBN code in Mathematica that incorporates (i) a modified time-temperature relation and the scale factor of the Universe, (ii) a modified electron neutrino distribution function (modifying the \pn conversions as mediated by weak interactions), and (iii) additional meson-driven \pn and nuclear dissociation rates; details may be found in Appendix~\ref{app:BBN-chain}.\footnote{This BBN code \texttt{BBNEasyALP} is publicly available at~\gitlink. The public version is identical to the one used in the paper with the only small approximation of considering only charged pions as injected mesons and considers them at rest. This has only a small impact on the parameter space for long ALP lifetimes and large ALP masses, but simplifies the code significantly.} For calculating the weak \pn rates, we first compute the bare rates in Born approximation $\Gamma_{\text{Born}}$, and then multiply them by the ratio $(\Gamma_{\text{corr}}/\Gamma_{\text{Born}})_{\text{SBBN}}$, incorporating the $\sim 2\%$ effect of radiative and nuclear structure corrections in the standard cosmological scenario. We take the latter and the default BBN nuclear rates from the PRIMAT BBN code~\cite{Pitrou:2018cgg}. We believe that our description of the $p\leftrightarrow n$ rates is accurate, as (i) there are no sizable neutrino spectral distortions, and (ii) our Standard Model results agree very well with the PRIMAT output, with differences well below observational errors on the primordial element abundances. Finally, we take the meson-driven rates from Refs.~\cite{Reno:1987qw,Pospelov:2010cw} (see also~\cite{Akita:2024ork}).

\textbf{BBN and CMB constraints.} To derive the BBN and CMB constraints, we use the most recent results from cosmological observations. For the deuterium abundance we use the value recommended by Particle Data Group~\cite{ParticleDataGroup:2024cfk} which is primarily based on  primarily on~\cite{Cooke:2017cwo}, and for the primordial helium-4 abundance we use the very recent and precise measurement from~\cite{Aver:2026dxv}:
\begin{subequations}\label{eq:BBN_data}
\begin{align}
        Y_{\rm P} &=0.2458 \pm 0.0013 \,,\\
   10^5\times  {\rm D/H}|_{\rm P} &= 2.547\pm 0.029\,.\label{eq:MeasuredD/H}
\end{align}
\end{subequations}
The $Y_{\rm P}$ measurement is in agreement with but significantly more precise and robust than previous analyses~\cite{Aver:2020fon,Valerdi:2019beb,Fernandez:2019hds,Kurichin:2021ppm,Hsyu:2020uqb,2021MNRAS.505.3624V,Aver:2021rwi}.

The ${\rm D/H}|_{\rm P}$ prediction depends strongly upon the baryon density of the Universe and is subject to uncertainties from nuclear reaction rates. There are several theoretical predictions for the deuterium abundance in the literature:
\begin{subequations}
\begin{align}
10^5  \,  {\rm D/H}|_{\rm P}^{\rm SM} &= 2.51\pm 0.07\,,~\cite{Pisanti:2020efz,Gariazzo:2021iiu}\\
10^5  \,  {\rm D/H}|_{\rm P}^{\rm SM} &= 2.48\pm 0.08\,,~\cite{Yeh:2020mgl,Yeh:2022heq}\\
10^5  \,  {\rm D/H}|_{\rm P}^{\rm SM} &= 2.44\pm 0.04\,, ~\cite{Pitrou:2018cgg,Pitrou:2020etk}\label{eq:DHPRIMAT}
\end{align}
\end{subequations}
where these predictions include the uncertainties from both nuclear reaction rates and the uncertainty in $\Omega_bh^2 = 0.02242\pm 0.00014$ as inferred by Planck~\cite{Planck:2018vyg}. Contrasting directly these expectations with the measured value in Eq.~\eqref{eq:MeasuredD/H} and adding the errors in quadrature, one would obtain the following allowed regions at $2\sigma$:
\begin{subequations}
\begin{align}
 {\rm D/H}|_{\rm P}/ [{\rm D/H}|_{\rm P}^{\rm SM}] &\in [-4.6\,,+7.5]\,\%\,,~\cite{Pisanti:2020efz,Gariazzo:2021iiu}\\
 {\rm D/H}|_{\rm P}/ [{\rm D/H}|_{\rm P}^{\rm SM}]  &\in [-4.1\,,+9.6]\,\%\,,~\cite{Yeh:2020mgl,Yeh:2022heq}\\
 {\rm D/H}|_{\rm P}/ [{\rm D/H}|_{\rm P}^{\rm SM}] &\in [+0.3\,,+8.4]\,\%\,, ~\cite{Pitrou:2018cgg,Pitrou:2020etk}
 \end{align}
\end{subequations}
Clearly, the first two agree well with the measured value in Eq.~\eqref{eq:MeasuredD/H} while the last one is $\sim 2\sigma$ lower. The approaches followed in these references to obtain the rates are all valid. By default, we used the rates from~\cite{Pitrou:2018cgg,Pitrou:2020etk} as reported in PRIMAT. Since the effects from our ALPs enter multiplicatively, to be maximally conservative, we will consider as the allowed $2\sigma$ region:
\begin{align}
     {\rm D/H}|_{\rm P}/ [{\rm D/H}|_{\rm P}^{\rm SM}] &\in [-4.6\,,+9.6]\,\%\,,
\end{align}
which, given that our SBBN calculation, predicts $10^5  \,  {\rm D/H}|_{\rm P}^{\rm SM} = 2.44 $ represents $10^5  \,  {\rm D/H}|_{\rm P}^{\rm SM} \in [2.31,\,2.67]$. 

We note that if new nuclear cross-section data or theoretical predictions reinforce the results of~\cite{Pitrou:2018cgg,Pitrou:2020etk} that the predicted deuterium abundance in the Standard Model is lower than the measured one, it would be a clear hint for BSM physics. In fact, the ALPs we consider can actually account for this, and we highlight this in Fig.~\ref{fig:results-Y-tau-plane} by showing the region 
\begin{equation}
{\rm D/H}|_{\rm P}/ [{\rm D/H}|_{\rm P}^{\rm SM}] \in [+2.2\,,+6.4]\,\%\,,
\label{eq:preferred-DH-Neff}
\end{equation}
where taking these nuclear rates at face value, the potential tension would be reduced to less than $1\sigma$. 

Contrasting the Helium-4 measurements to our predictions is much simpler, as the theoretical uncertainty in its calculation is negligible compared to~\ref{eq:BBN_data}, leading to $Y_{\rm P}^{\rm SM} = 0.247$. Following the same procedure allows us to define the $2\sigma$ allowed parameter space as:
\begin{align}
     Y_{\rm P}/Y_{\rm P}^{\rm SM}\in [-1.4\,,0.7\,]\,\%\,,
\end{align}
corresponding to 
$Y_{\rm P} \in [0.2433\,,0.2483]$.

In the context of CMB observations, we use the latest and most precise measurement of \neff from the combination of Planck~\cite{Planck:2018vyg}, ACT~\cite{ACT:2025fju,ACT:2025tim}, and SPT~\cite{SPT-3G:2025bzu} data that has been derived this year:
\begin{align}\label{eq:Neff_inference}
    \neff = 2.81\pm 0.12 \,.
\end{align}
We note that $\neff$ CMB inferences are correlated with the matter density and $H_0$, and that these CMB-inferred parameters are in a $\sim 3\sigma$ tension with DESI BAO data~\cite{DESI:2025zgx}. Since this tension would only tend to increase the inferred value of \neff, whereas in the ALP model we consider \neff can only be smaller than the Standard Model prediction, our treatment is conservative. We take Eq.~\eqref{eq:Neff_inference} at face value,  which implies an allowed interval $\neff \in [2.58,3.05]$ at 95\% CL. In practice, we will adopt $2.58$ as the relevant bound; however, in some figures we will also indicate in red the $1\sigma$ region corresponding to~\eqref{eq:Neff_inference}, to emphasize that for those ALPs we consider the resulting value of this cosmological parameter would fall into $1\sigma$ concordance.

\section{Results}
\label{sec:results}

\begin{figure}[t!]
    \centering
    \includegraphics[width=\linewidth]{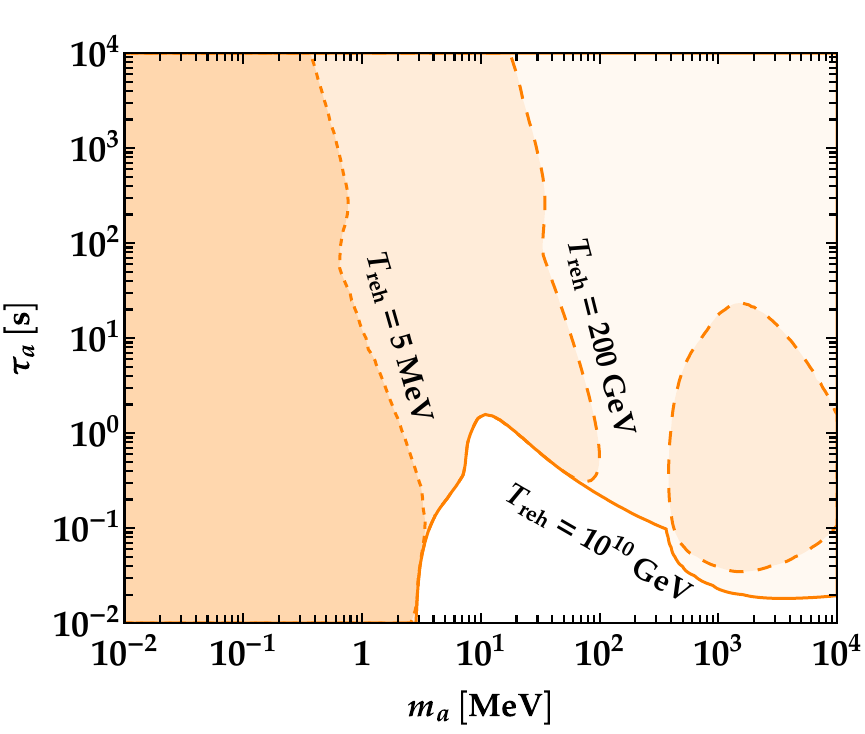}
    \caption{The cosmological constraints on ALPs interacting with photons shown for several values of the reheating temperature $\Treh = 5\mev, 200\text{ GeV}, 10^{10}\text{ GeV}${, with the boundaries of the constraints indicated by the dashed lines}. The bounds combine constraints from \neff, \yp, and \dH. The island for $\Treh = 200\gev$ results from the impact of the rare decays of the ALP into {metastable mesons $\pi^{\pm},K^{\pm},K_{L}$. The injected $\pi^\pm$ trigger fast $p\leftrightarrow n$ conversions, so the BBN impact can persist even when the ALP abundance is reduced by late reheating (see Sec.~\ref{sec:BBN-chain-details} for more details)}. We refer to Fig.~\ref{fig:comparison-robust} in appendix where we show limits for $\Treh = 10\mev, 1\text{ GeV}, 10^3\text{ GeV}, 10^6\text{ GeV}$.}
    \label{fig:results-ALPs-reheating-temperature}
\end{figure}

\begin{figure*}[t!]
    \centering
    \includegraphics[width=0.5\linewidth]{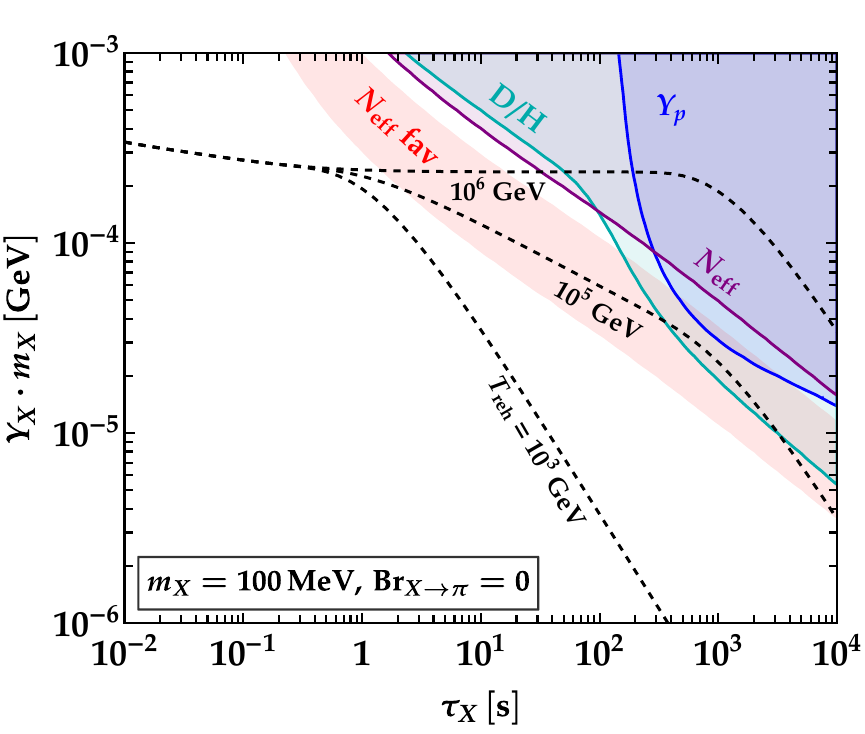}~\includegraphics[width=0.5\linewidth]{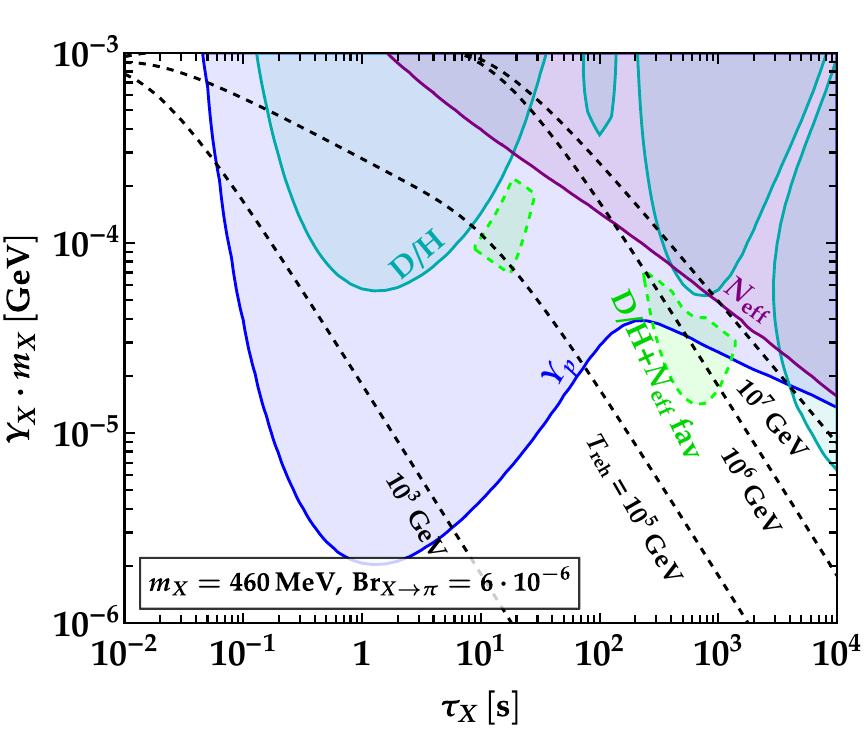}\\  \includegraphics[width=0.5\linewidth]{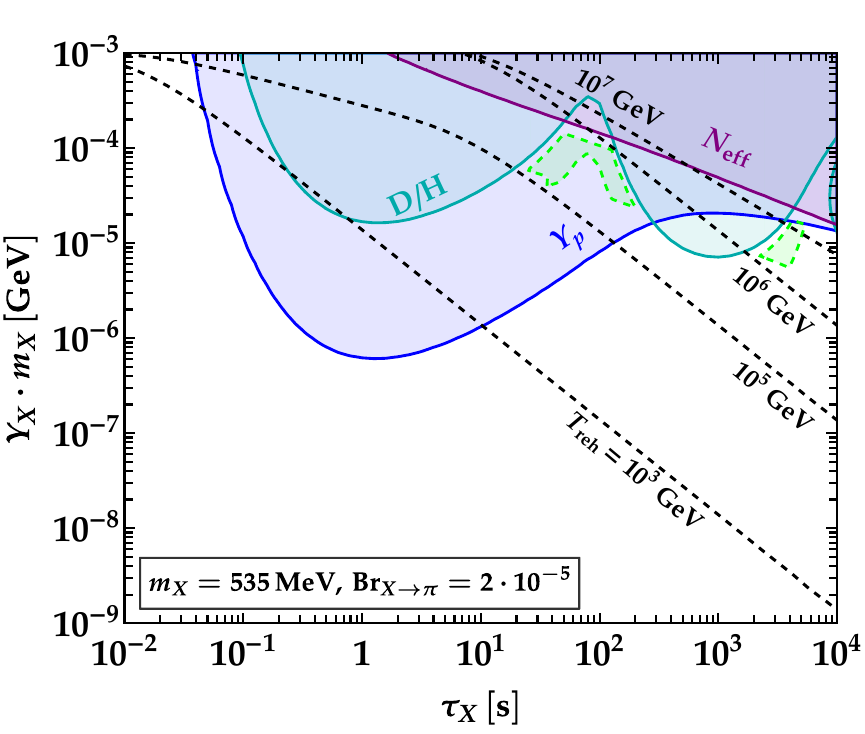}~\includegraphics[width=0.5\linewidth]{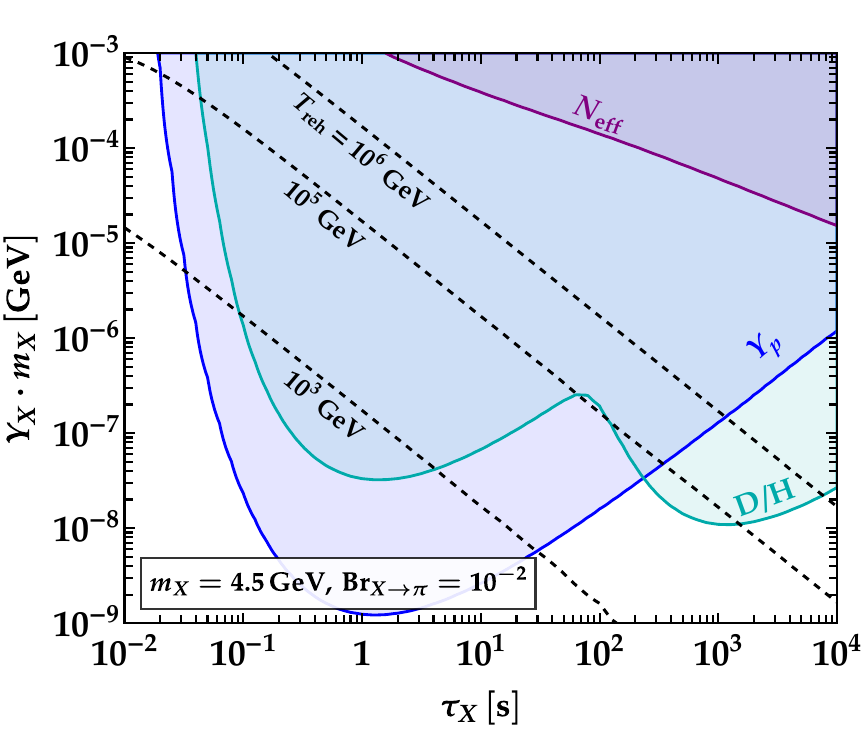}~
    \caption{The cosmological constraints on a particle $X$ that primarily decays into $\gamma \gamma$ and with abundance $Y_{X} = n_{X}/s$ at $T = 20\mev$ (assuming that it is fully decoupled), mass $m_{X}$, and lifetime $\tau_{X}$, in the plane $Y_{X}\cdot m_{X}$ vs $\tau_{X}$. Several values of the mean number of pions per $X$ decay and mass are shown: $m_{X} = 100\mev, \text{Br}_{X\to \pi}=0$ (top left), $460\mev,6\cdot 10^{-6},$ (top right), $535\mev,2\cdot 10^{-5}$ (bottom left), and $4.5\gev,10^{-2}$ (bottom right) corresponding to the $\text{Br}_{X\to \pi}$ expected from our calculation in Sec.~\ref{sec:methodology}. The green domains correspond to the region preferred by the combination of \dH and \neff measurements, see the discussion around Eq.~\eqref{eq:preferred-DH-Neff}. The plot may be mapped onto the ALP parameter space in scenarios with arbitrary cosmological setups; we indicate this map by showing the iso-contours of the ALP abundance with the mass $m_{a} = m_{X}$ for fixed reheating temperatures $\Treh${, which we denote by the dashed black lines.} }
    \label{fig:results-Y-tau-plane}
\end{figure*}

\textbf{Parameter space for very high reheating.} In {the left panel of} Fig.~\ref{fig:ALP-results}, we show our constraints on the ALP parameter space considering a very high reheating temperature $\Treh \ge  10^{10}\gev$. The behavior of the various limits is easy to understand. The first thing to note is that the reheating temperature is so high that the ALPs were in thermal equilibrium for a large portion of the parameter space we explore. This means that ALPs will arrive at BBN times with a number density comparable to that of photons. In this context, as the ALP mass increases, its energy density will be larger, which explains why the constraints from all probes become stronger for $m_a > 10\,{\rm MeV}$. One can also see the effect of the two-pion threshold: above $m_{a} \simeq 2m_{\pi}$, the \yp constraint can reach $\tau_a \sim 0.02\,{\rm s}$, comparable with the constraints on the particles which dominantly decay into hadrons~\cite{Kawasaki:2017bqm,Fradette:2018hhl,Boyarsky:2020dzc,Jung:2025dyo}. At $m_a\lesssim 2m_{\pi^\pm}$, the bounds are essentially dictated by the requirement that the energy density injected in photons is not more than $\sim 20\%$ of the neutrino one. The \yp limits are weaker in this region because \yp is mainly sensitive to the primordial neutron abundance, which is set at the plasma temperature $T_\gamma \simeq 0.7\mev$, and the injection happens substantially later. We refer to Appendix~\ref{app:some-results} where we show the iso-contours for each of these cosmological observables. 

\textbf{Parameter space for various reheating temperatures.} In Fig.~\ref{fig:results-ALPs-reheating-temperature}, we show the resulting combined BBN+CMB limits for various reheating temperatures: $\Treh = 5\mev$ (lowest possible reheating temperature ~\cite{Hasegawa:2019jsa,Barbieri:2025moq}), $\Treh = 200\gev$ (slightly above sphaleron freeze-out~\cite{DOnofrio:2014rug}), $\Treh = 10^{10}\gev$ (a temperature where thermal leptogenesis could still be effective~\cite{Giudice:2003jh}). The line with $\Treh = 5\mev$ leads to the irreducible bound~\cite{Langhoff:2022bij}. Interestingly, for the domain $10 \,{\rm GeV}\lesssim  \Treh \lesssim 1\,{\rm TeV}$, we see an island around $m_a \in [0.5 \,,10]\gev$ and $\tau_a \in [0.1 \,,10]\,{\rm s}$, coming primarily from \yp bounds. This is a new result from our study, stemming from the fact that we included the effects of rare hadronic decays in the BBN evolution {(recall the discussion in Sec.~\ref{sec:BBN-chain-details})}.   

\textbf{Parameter space for generic relics decaying into photons.} In Fig.~\ref{fig:results-Y-tau-plane}, we cast our results in terms of the mass ($m_X$), lifetime ($\tau_X$), and yield $Y_X$ at $T = 20\mev$ of a generic relic $X$ that decays into two photons, having also a sub-dominant decay mode into charged pions. To make it interpretable as the ALP parameter space, for each mass $m_{X}$, we aligned the pionic branching ratio according to our calculation in Section~\ref{sec:methodology}; we also show iso-contours of the abundance that would be obtained if this particle were an ALP, given a reheating temperature. 

The panels clearly show the important impact of considering the rare decays of these particles into hadrons{. In particular, the top left panel shows that without hadronic decays, the $Y_{p}$, ${\rm D/H}$ constraints are at the level of the \neff region, because they all are controlled by the ALP energy density (recall a discussion in Sec.~\ref{sec:cosmoeffects}). Including the hadronic decays makes the BBN constraints only weakly sensitive to the ALP energy density. Even for tiny $\text{Br}_{X\to\pi}\sim 10^{-2}$, the bounds may get strengthened by six orders of magnitude, depending upon mass and lifetime.} These panels also allow us to easily understand the island found in Fig.~\ref{fig:results-ALPs-reheating-temperature} for $\Treh = 200\gev$. 

In Fig.~\ref{fig:results-Y-tau-plane}, the various colored lines have the same meaning as the left panel of Fig.~\ref{fig:ALP-results}, but here we show an additional region in green that could reconcile \textit{simultaneously} two different small cosmological tensions: 1) it would lead to a value of $\neff = 2.81\pm0.12$ and hence agree with the central value of the latest CMB measurements, and 2) it would also correspond to a deuterium abundance which is $2-6$\% larger than the Standard Model one and which would be relevant if a consensus emerges that the SM deuterium prediction is smaller than that measured astronomically, see Sec.~\ref{sec:methodology}. While clearly these hints are not statistically significant and could very well go away, the figure allows us to identify the relevant parameter space where they would be addressed: $m_a \sim 500\mev$, ${\rm Br}(a\to \gamma \pi^+\pi^-) \sim 10^{-5}$, with $\tau_a \sim 300-5000\,{\rm s}$ and with an abundance corresponding to that generated for an ALP with a reheating temperature $\Treh \sim (5\times10^{5}-10^7)\gev$. { It is interesting to note that this region of parameter space is not too far away from the $Y_{\rm P}$ constraint which highlights that an even more precise measurement of the $Y_{\rm P}$ abundance can help rule in or out this region.}

\textbf{Comparison with previous works.} Our results can first be compared with the classic analysis of photophilic ALPs in \cite{Cadamuro:2011fd}, which effectively assumed $\Treh = \infty$. At the qualitative level, we recover the same structure of the excluded region, in particular the sharp strengthening of the $Y_{\rm P}$ constraint once $m_a \simeq 2m_{\pi^\pm}$. Quantitatively, however, the allowed domain looks different because~\cite{Cadamuro:2011fd} used older determinations of the primordial abundances and, in practice, imposed only a conservative lower limit on ${\rm D/H}|_{\rm P}$ and an upper limit on $Y_{\rm P}$. Our bounds are therefore more stringent, since with present data one must also exclude downward shifts of $Y_{\rm P}$ and upward shifts of ${\rm D/H}|_{\rm P}$. 

A second relevant comparison is with~\cite{Depta:2020wmr}, which already explored how the constraints for several $\Treh$. The main difference is that~\cite{Depta:2020wmr} did not include the rare but cosmologically important mesonic decay modes of photophilic ALPs. As a consequence, for $\Treh = \infty$, their BBN exclusions in the range $m_a \gtrsim 500\mev$ are typically weaker by about one order of magnitude in terms of lifetime, since only the electromagnetic decays were considered. Moreover, in the intermediate window $10\,{\rm GeV}\lesssim \Treh \lesssim  1\,{\rm TeV}$, neglecting meson decays removes the broad excluded islands that we find, because in that case the only handle is the ALP energy density, and this rapidly decreases once $\Treh$ is lowered. 

In Appendix~\ref{app:previous-works-comparison}, we provide a dedicated comparison with each of these references.  

\textbf{Implications for generic ALPs.} As a concluding remark, let us comment on the case of generic ALPs, having various masses and coupling patterns, including the interactions with $SU(2)_{L}$ fields, gluons, and fermions. The $SU(2)_{L}$ coupling may be treated very similarly to the photonic ALP. In particular, the dominant decay of such ALPs is still $a\to \gamma\gamma$. In the case of hadronic couplings, the situation is more complicated; the main reason is that hadronic decays dominate. First, hadronic decays for GeV-mass ALPs suffer from sizeable theoretical uncertainties (see Ref.~\cite{Ovchynnikov:2025gpx}), which propagate into the lifetime-coupling relation and hadronic multiplicities. Second, the metastable hadronic decay products of the ALPs, $\pi^{\pm},K^{\pm},K_{L}$, undergo various interaction processes, which influence the distribution of energy between the EM and neutrino sectors, and have to be accurately captured to properly understand the dynamics of the Universe~\cite{Akita:2024nam,Akita:2024ork}. Finally, if decaying, such mesons inject non-thermal neutrinos; as a result, the evolution of neutrinos (important to obtain \neff and \yp constraints) must be traced using the unintegrated neutrino Boltzmann equation~\cite{Ovchynnikov:2024xyd,Ihnatenko:2025kew}.

\section{Conclusions}
\label{sec:conclusions}

In this paper, we have thoroughly investigated the impact of axion-like particles (ALPs) coupled to a pair of photons on the evolution of the Universe. We focused on ALPs with lifetimes $\tau_{a}\lesssim 10^{4}\s$ and masses $m_a\lesssim 10\,{\rm GeV}$ and their implications for BBN and the CMB. We have systematically incorporated all relevant elements of the ALPs' influence in the Universe, including: i) expansion, ii) modification to weak $\pn$ conversion rates, and, critically, iii) the meson-driven \pn conversion and nuclear dissociation processes. {Our Mathematica BBN code \texttt{BBNEasyALP} is publicly available at~\gitlink}.

We have used precision data on $N_{\rm eff}$, helium $Y_{\rm P}$, and deuterium ${\rm D/H}|_{\rm P}$ to set limits on the lifetime and mass of ALPs as a function of the reheating temperature. Our main results are summarized in Fig.~\ref{fig:ALP-results}{. In particular, in its right panel, we show the combination of the cosmological constraints derived in this study together with the existing bounds from astrophysical observations and laboratory searches. O}ne can see that these bounds are in some regions the most constraining ones, while in others, they are complementary to astrophysical and laboratory limits.

Critically, we have also considered how the bounds weaken when the reheating temperature of the Universe is low. Our results are shown in Fig.~\ref{fig:results-ALPs-reheating-temperature}. In this context, we have shown that taking into account the effect of rare ALP decays into mesons is key. In particular, it has allowed us to set limits on some regions of parameter space which were thought to be cosmologically viable.  

While across our results we mainly show excluded regions, in Fig.~\ref{fig:results-Y-tau-plane} we have identified the parameter space where ALPs coupled to a pair of photons could actually ameliorate simultaneously two small tensions: the currently slightly smaller than 3 $N_{\rm eff}$ CMB measurement, see Eq.~\eqref{eq:Neff_inference}, and the higher measured deuterium abundance as compared to the SM prediction obtained using some sets of nuclear reaction rates, see Eq.~\eqref{eq:DHPRIMAT}. 

ALPs are a generic prediction of low-energy realizations of string theory. While there are no firm predictions on the actual spectrum and couplings for them, some of these axions may end up having masses and lifetimes in the window we focused on in our study. We have presented results in a model-independent fashion (see Fig.~\ref{fig:results-Y-tau-plane}), {and made our codes publicly available}. It is our hope that this will allow interested researchers to find exact cosmological limits for their models, but also allow for generalizations and expansions. The latter includes considering longer lifetimes, the interplay with photo-disassociation, as well as going to higher ALP masses and considering various ALP coupling patterns.

\section*{Acknowledgements}

We have used the nuclear reaction rates as tabulated in PRIMAT~\cite{Pitrou:2018cgg}, and we are grateful to their authors for their online availability. MO received support from the European Union's Horizon Europe research and innovation programme under the Marie Sklodowska-Curie grant agreement No~101204216. 

\newpage 
\appendix
\onecolumngrid

\section*{Supplemental Material}\label{app:appendices}

In this Supplemental Material, we discuss technical details that are necessary to understand the impact of ALPs on the Universe.

It is organized as follows. In Sec.~\ref{app:alp-production}, we formulate the Boltzmann equation governing the evolution of the ALP population, and derive the rates of the ALP production processes in the Early Universe. Sec.~\ref{app:alp-decay} is devoted to discussions of the ALP decay widths, including the hadronic modes $a\to \gamma +\text{hadrons}$. Sec.~\ref{app:thermodynamics} discusses the approach to derive the thermodynamics of the Universe modified by the ALPs. Finally, Sec.~\ref{app:BBN-chain} summarizes the modification of the \pn and nuclear chain in the presence of the ALP decay products. Finally, Sec.~\ref{app:previous-works-comparison} compares our results with the previous works.

\section{ALP production}
\label{app:alp-production}

To understand the production of ALPs depending on the reheating temperature, we need to carefully calculate the production rates as a function of the ALP mass and temperature in the Universe, taking into account the various particle species participating in the production. This section is devoted to the discussion of the evolution of the ALP population, which we parameterize in terms of the ALP abundance 
\begin{equation}
\mathcal{Y}_{a}(T) \equiv \frac{n_{a}}{s},
\end{equation}
where $n_{a}$ is the ALP number density and $s = 2\pi^{2} g_{*,s}T^{3}/45$ is the entropy density of the Universe. 

Let us first summarize the main approximations used in this section.
\begin{enumerate}
\item We consider $T\gg 1\mev$ (concretely, $T  >T_{\text{split}} = 20\mev$), where the ALPs do not dominate the energy density of the Universe. This is the case even for the heaviest ALP we consider ($m_a = 10\gev$, assuming it has a long lifetime and is produced with a relativistic density at $ T \gg m_a$). As such, we consider that the Universe is dominated by the particles in the Standard Model and use the effective degrees of freedom contributing to energy density and entropy as in the Standard Cosmological Scenario from~\cite{Laine:2006cp,Laine:2015kra}. This is, $H = 1.66 \sqrt{g_\star} T^2/m_{\rm pl}$ and $s = 2\pi^2 g_{s\star} T^3/45$.
\item We also assume that all the particles producing the ALPs are in thermal equilibrium, given the strength of the Standard Model interactions. In particular, we consider that they are described by Bose-Einstein or Fermi-Dirac distribution functions with a common temperature $T$ and negligible chemical potentials. 
\item Finally, we neglect the contribution of the ALP interaction with the $Z$ boson to the production. It may emerge from the $U(1)_{Y}\otimes SU(2)_{L}$ completion of the effective ALP interaction 
\begin{equation}
g_{a\gamma\gamma}aF_{\mu\nu}\tilde{F}^{\mu\nu}\to \frac{g_{a\gamma\gamma}}{\cos^{2}(\theta_{W})}B_{\mu\nu}\tilde{B}^{\mu\nu},
\end{equation}
with $B$ being the $U(1)_{Y}$ hypercharge field and $\theta_{W}$ the Weinberg's angle. We do not expect it to contribute significantly to the ALP production, as the $aZ\gamma$ and $aZZ$ operators resulting from this approximation would, contributing in a similar fashion to the $a\gamma\gamma$ operators, be simultaneously suppressed by the powers $\tan(\theta_{W}), \tan^{2}(\theta_{W})$ correspondingly.
\end{enumerate}

The equation governing the evolution of $\mathcal{Y}_{a}$ has the form
\begin{equation}
    \frac{d\mathcal{Y}_{a}}{dt} = \frac{1}{s}\sum_{i}\big(n_{i}n_{\gamma}\langle \sigma v\rangle_{i\gamma\to ia} - n_{a}n_{i}\langle\sigma v\rangle_{ia\to i\gamma}\big)+\frac{1}{s}\left(n_{\gamma}^{2}\langle \sigma v\rangle_{\gamma+\gamma\to a} - n_{a}\langle\Gamma\rangle_{a\to \gamma+\gamma}\right)\,.
    \label{eq:ALP-production}
\end{equation}
Here, $i$ sums over species participating in the $2\to 2$ scattering $i+\gamma\to i+a$ (called the Primakoff process). The second term is the photon fusion and the backward ALP decay. $\langle \sigma v\rangle$ is the cross-section-times-velocity averaged over the distributions of the incoming particles, and $\langle\Gamma\rangle_{a\to \gamma+\gamma}$ is the ALP width averaged over the ALP distribution. 

Equation~\eqref{eq:ALP-production} can be simplified using the detailed balance principle:
\begin{equation}
    n_{i}^{\text{eq}}n_{\gamma}^{\text{eq}}\langle \sigma v\rangle_{i\gamma\to ia} = n_{a}^{\text{eq}}n_{i}^{\text{eq}}\langle\sigma v\rangle_{ia\to i\gamma}, \quad (n_{\gamma}^{\text{eq}})^{2}\langle \sigma v\rangle_{\gamma+\gamma\to a} = n_{a}^{\text{eq}}\langle \Gamma_{a\to \gamma+\gamma}\rangle\,,
\end{equation}
Thus, we have
\begin{align}
    \frac{d\mathcal{Y}_{a}}{dt} &= \frac{1}{s}\sum_{i}n_{i}^{\text{eq}}n_{\gamma}^{\text{eq}}\langle \sigma v\rangle_{i\gamma\to ia}\left(1-\frac{n_{a}}{n_{a}^{\text{eq}}}\right) + \frac{1}{s}(n_{\gamma}^{\text{eq}})^{2}\langle \sigma v\rangle_{\gamma+\gamma\to a}\left(1-\frac{n_{a}}{n_{a}^{\text{eq}}} \right) \nonumber \\
    &=\sum_{i}\frac{n_{i}^{\text{eq}}n_{\gamma}^{\text{eq}}}{n_{a}^{\text{eq}}}\langle \sigma v\rangle_{i\gamma\to ia}(\mathcal{Y}_{a}^{\text{eq}} - \mathcal{Y}_{a}) + \frac{(n_{\gamma}^{\text{eq}})^{2}}{n_{a}^{\text{eq}}}\langle\sigma v\rangle_{\gamma\gamma\to a}(\mathcal{Y}_{a}^{\text{eq}} - \mathcal{Y}_{a}) \nonumber \\
    &\equiv \Gamma_{a}\cdot (\mathcal{Y}_{a}^{\text{eq}} - \mathcal{Y}_{a})\,,
    \label{eq:alp-abundance-equation}
\end{align}
where we have introduced the ALP production rate $\Gamma_{a}$, given by
\begin{equation}
    \Gamma_{a}(T) \equiv  \sum_{i}\frac{n_{i}^{\text{eq}}n_{\gamma}^{\text{eq}}}{n_{a}^{\text{eq}}}\langle \sigma v\rangle_{i\gamma\to ia} + \frac{(n_{\gamma}^{\text{eq}})^{2}}{n_{a}^{\text{eq}}}\langle\sigma v\rangle_{\gamma\gamma\to a}\equiv \Gamma_{a}^{\text{Prim}} + \Gamma_{a}^{\text{fusion}}
    \label{eq:ALP-total-production-rate}
\end{equation}
We integrate equation~\eqref{eq:alp-abundance-equation} starting at the reheating temperature $T = \Treh$ with $Y_a = 0$ until $T = T_{\text{split}} = 20\mev$ $T\in (T_{\text{split}},\Treh)$, where $\Treh$ is the reheating temperature. Next, we define 
\begin{equation}
Y_{a} \equiv \mathcal{Y}_{a}(T_{\text{split}}),
\label{eq:ALP-abundance-final}
\end{equation}
which we will use in Sec.~\ref{app:thermodynamics} as the initial condition for the ALP population when considering the temperatures $T<T_{\text{split}}$. 

\textbf{ALP production rates:} We consider the production from all charged SM fermions: $i = e,\mu,\tau$ and quarks $q = u,d,s,c,b,t$. At temperatures above the scale of the electroweak phase transition (assumed to be $\Lambda_{\text{EW}} = m_{t}$), we smoothly set particles' masses to zero. Also, we smoothly turn off the contributions of the quarks below the temperature $T=280\mev$, corresponding to the QCD crossover.  Namely, we suppress the quark-driven terms in the plasmon mass and the rates by a smooth exponential factor that $\to 1$ at $T > 280\mev$ and sharply reduces to zero at $T<280\mev$.
 
\begin{figure}[h!]
    \centering
    \includegraphics[width=0.5\linewidth]{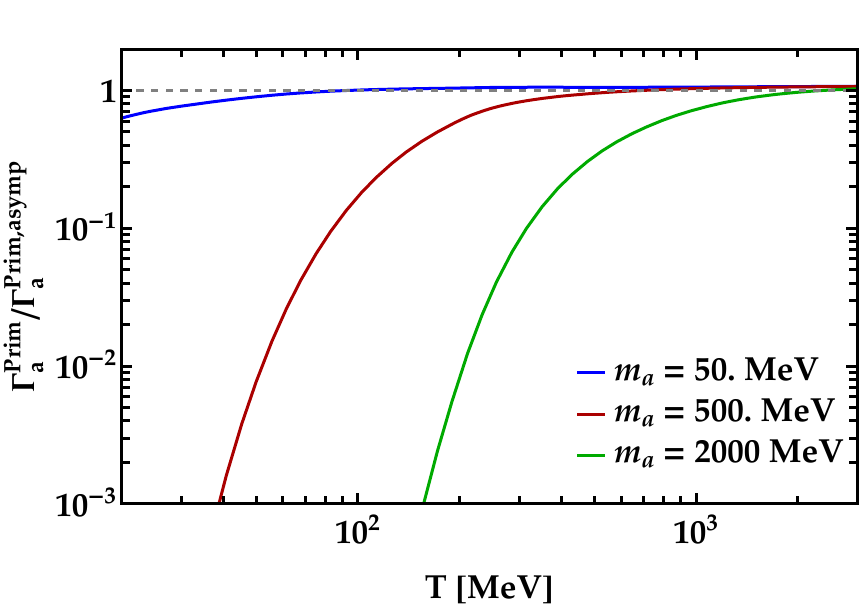}\includegraphics[width=0.5\linewidth]{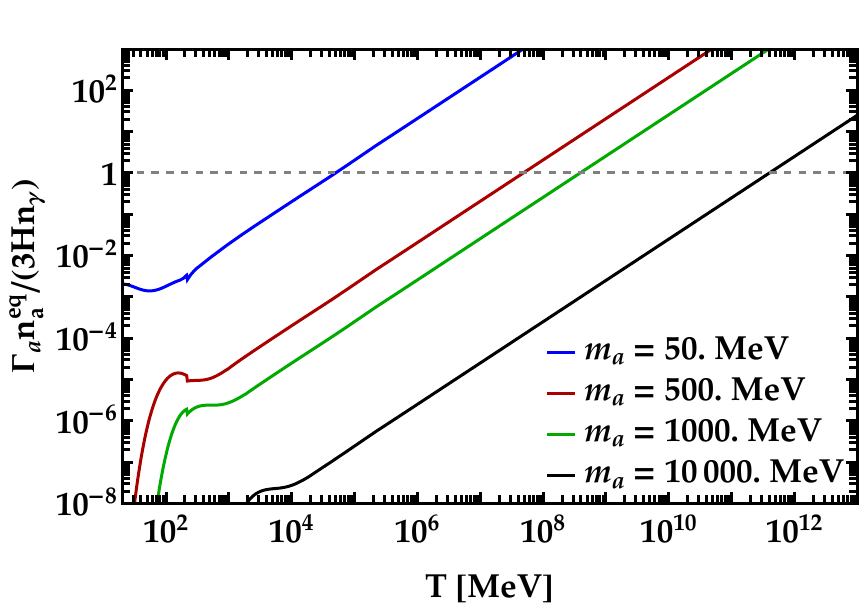}
    \caption{\textit{Left panel}: the ratio $\Gamma_{a}^{\text{Prim}}/\Gamma_{a}^{\text{Prim,asymp}}$, where $\Gamma^{\text{Prim}}_{a}$ is the ALP production rate in the Primakoff process given by Eq.~\eqref{eq:ALP-total-production-rate}, whereas $\Gamma_{a}^{\text{prim,asymp}}$ is the asymptotics given by Eq.~\eqref{eq:ALP-production-rate-massless-asymptotics}. Several ALP masses are considered. \textit{Right panel}: the ratio $\Gamma_{a}\cdot (n_{a}^{\text{eq}}/n_{\gamma})/3H$, which controls whether the ALPs may enter thermal equilibrium at the given temperature $T$. The value of the ALP coupling is fixed by requiring the lifetime to be $\tau_{a} = 0.1\s$. Independent of the ALP mass, the ratio is the highest at large temperatures (being driven by the Primakoff process), then smoothly decreases, reaches a minimum, and starts increasing (being driven by the inverse ALP decay rate), asymptotically reaching $\tau_{a}^{-1}$. {The dashed gray lines indicate the equality between the rates, which we interpret as the parameter space where the ALPs become in thermal equilibrium.}}
    \label{fig:ALP-production-rates}
\end{figure}

Using the resulting thermally averaged cross-section $\langle \sigma v\rangle_{i\gamma\to ia}$ as calculated in Sec.~\ref{app:production-cross-section-calculation}, we show the behavior of the ALP production rates for several choices of the ALP mass in Fig.~\ref{fig:ALP-production-rates}. The left panel shows a cross-check of our approach -- reproducing the asymptotic result $m_{a}\ll T$ presented in~\cite{Bolz:2000fu} (Eq.~(31)), that is
\begin{equation}
    \Gamma_{a}^{\text{asymp}} = \frac{1}{n_{a}^{\text{eq}}}\frac{\sum_{i}n_{i}Q_{i}^{2}}{n_{e}}\frac{g_{a\gamma\gamma}^{2}\alpha_{\text{EM}}\zeta(3)T^{6}}{12\pi^{2}}\left( \log\left( \frac{T^{2}}{m_{\gamma}^{2}}\right)+0.8194\right)\,,
    \label{eq:ALP-production-rate-massless-asymptotics}
\end{equation}
The right panel shows the ratio $\Gamma_{a}/3H$, which defines whether ALPs may reach thermal equilibrium at the given temperature $T$. Using it, we may qualitatively conclude on the value of the reheating temperature \Treh for which the ALP abundance for the given mass and lifetime quickly decreases. 

\begin{figure}[h!]
    \centering
    \includegraphics[width=0.5\linewidth]{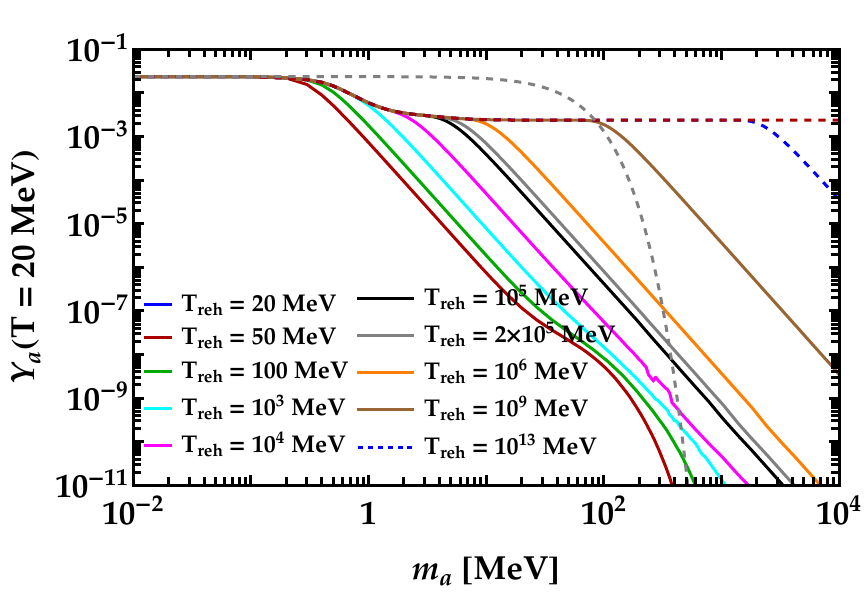}~\includegraphics[width=0.5\linewidth]{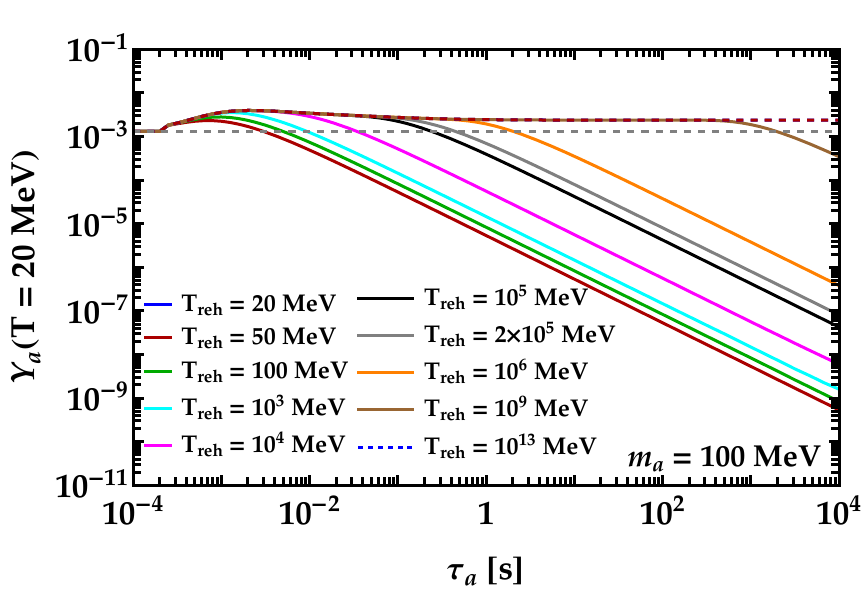}
    \caption{The ALP mass (left panel) and ALP lifetime (right panel) dependence of the ALP abundance $Y_{a} \equiv (n_{a}/s)|(T = 20\mev)$, assuming different values of the late reheating temperature $\Treh$. The gray dashed line shows the abundance if assuming the ALPs were in thermal equilibrium.}
    \label{fig:abundances-behavior}
\end{figure}

Fig.~\ref{fig:abundances-behavior} shows the behavior of $Y_{a}$ with the ALP mass (left panel) or the ALP lifetime $\tau_{a}$ (right panel), for different values of the late reheating temperature $\Treh$. As we see, the light ALPs with mass $m_{a}\lesssim 10\mev$ couple significantly strongly to remain in equilibrium at these temperatures. Once mass or lifetime increases, the ALPs decouple earlier and earlier. This is because for the given lifetime, the ALP coupling $g_{a\gamma\gamma}$ scales as $g_{a\gamma\gamma}\propto \sqrt{1/(\tau_{a}m_{a}^{3})}$ (we used Eq.~\eqref{eq:ALP-dominant-width}). As a result, heavier ALPs decouple at larger temperatures. However, assuming the absence of late reheating, for the masses of interest $m_{a}<10\text{ GeV}$, the ALPs were in equilibrium at least for some period of time. If decoupling while being ultrarelativistic, their abundance is given by $Y_{a}\sim 10^{-3}$. Decreasing $\Treh$ leads to scenarios when the ALPs never entered equilibrium, i.e., have been produced via the freeze-in mechanism. In this regime, the ALP abundance scales as $Y_{a}\propto \Treh/(\tau_{a}m_{a}^{3})$, which follows from the asymptotics~\eqref{eq:ALP-production-rate-massless-asymptotics}.

\subsection{Averaged cross-section calculations}
\label{app:production-cross-section-calculation}
We calculate the thermally averaged cross section, $\langle \sigma v\rangle_{i\gamma\to ia}$. following the approach of Ref.~\cite{Edsjo:1997bg}. By definition:
\begin{equation}
    \sum_{i}n_{i}n_{\gamma}\langle \sigma v\rangle_{i\gamma\to ia} \equiv  \sum_{i}n_{i}n_{\gamma}\int W_{i\gamma}f_{i}(E_{i})f_{\gamma}(E_{\gamma})d\Phi_{i\gamma}\,,
    \label{eq:integral-generic}
\end{equation}
Here, $f_{i,\gamma}$ is the distribution function of the $i$ fermion and the photon; $n_{\alpha}$ is the number density:
\begin{equation}
n_{\alpha} \equiv N_{\alpha}\cdot g_{\alpha} \cdot \int \frac{d^{3}\mathbf{p}}{(2\pi)^{3}}f_{\alpha}(p,T),
\end{equation}
with $N_{i} = 3$ for quarks and $N_{i} = 1$ for leptons and photons, $g_{i} = 4, g_{\gamma} = 2$ the number of helicity and charge degrees of freedom. $d\Phi_{i\gamma}$ is the phase space of the incoming particles,
\begin{equation}
    d\Phi_{i\gamma} \equiv \frac{N_{i}g_{i}d^{3}\mathbf{p}_{i}}{(2\pi)^{3}2E_{i}}\frac{g_{\gamma}d^{3}\mathbf{p}_{\gamma}}{(2\pi)^{3}2E_{\gamma}}, \quad g_{i} = 4, \quad g_{\gamma} = 2,
\end{equation}
finally,
\begin{equation}
    W_{i\gamma} = \int \overline{|\mathcal{M}|_{i\gamma\to ia}^{2}}(2\pi)^{4}\delta^{4}(p_{i}+p_{\gamma} - p_{i}' - p_{a})\frac{d^{3}\mathbf{p}_{a}}{(2\pi)^{3}2E_{a}}\frac{d^{3}\mathbf{p}_{i}'}{(2\pi)^{3}2E_{i}'},
\end{equation}
with $\overline{|\mathcal{M}|_{i\gamma\to ia}^{2}}$ being the squared matrix element averaged over the polarizations of incoming particles.

Approximating $f_{i},f_{\gamma}$ by Maxwell-Boltzmann distributions and switching to the integration variables $E_{\pm} = (E_{i}\pm E_{\gamma}), s = (p_{i}+p_{\gamma})^{2}$, the integral~\eqref{eq:integral-generic} may be reduced to
\begin{equation}
 \sum_{i}n_{i}n_{\gamma}\langle \sigma v\rangle_{i\gamma\to ia}  \approx \frac{T}{32\pi^{4}} \sum_{i}\int \limits_{s_{\text{min}}}^{\infty}d s\ g_{i}g_{\gamma}p_{i\gamma}^{\text{CM}}(s)W_{i\gamma}(s)K_{1}\left(\frac{\sqrt{s}}{T}\right), 
\end{equation}
with $s_{\text{min}} = (m_{i}+m_{a})^{2}$, $p_{i\gamma}^{\text{CM}} = (s-m_{i}^{2})/2\sqrt{s}$, and $K_{1}$ the modified Bessel function of the second kind.

The remaining ingredient is $W_{i\gamma}$. It is given by
\begin{equation}
    W_{i\gamma}(s) = \frac{p_{a}^{\text{CM}}}{8\pi\sqrt{s}}\int d\cos(\theta) \overline{|\mathcal{M}|_{i\gamma\to ia}^{2}}(s,\cos(\theta)),
\end{equation}
where $\cos(\theta)$ is the center-of-mass (CM) scattering angle, and 
\begin{equation}
p_{a}^{\text{CM}} =\frac{\sqrt{s-(m_{a}-m_{i})^{2})(s-(m_{a}+m_{i})^{2})}}{2\sqrt{s}} \,,
\end{equation}
is the ALP momentum in the CM frame.

The squared matrix element has the form
\begin{equation}
    \overline{|\mathcal{M}|_{i\gamma\to ia}^{2}} = \frac{2}{g_{i}g_{\gamma}}\sum_{\text{polarizations}}\left|\mathcal{M}\right|^{2}, \quad \mathcal{M} = \frac{g_{a\gamma\gamma}eQ_{i}}{(p_{\gamma}-p_{a})^{2}-m_{\gamma}^{2}}\varepsilon^{\mu \nu\alpha\beta}\epsilon_{\mu}(p_{\gamma})(p_{\gamma})_{\nu}(p_{\gamma}-p_{a})_{\alpha}\bar{u}(p_{i}')\gamma_{\beta}u(p_{i})\,.
\end{equation}
Here, a factor of $g_{i}/2$ is the number of helicity degrees of freedom of the charged species $i^{\pm}$, $e = \sqrt{4\pi\alpha_{\text{EM}}}$ is the EM constant, $Q_{i}$ is the electric charge of the particle $i$ in the units of the electron charge, $\epsilon_{\mu}$ is the polarization vector of the incoming photon, while $m_{\gamma}$ is the thermal plasmon mass, given by~\cite{Fornengo:1997wa}
\begin{equation}
m_\gamma^2(T_\gamma) = \sum_{X}2g_{X}\alpha_{\rm EM}T^2/\pi \,\int_0^{\rm \infty} dx \,\frac{x^2}{(1+e^{\sqrt{x^2+(m_{X}/T)^2}})\sqrt{x^2+(m_{X}/T)^2}},
\label{eq:mplasmon}
\end{equation}
with the sum running over all SM fermions $X = e,\mu,\tau,\dots$, and $g_{X}$ being the number of degrees of freedom (including spin, charge, and colors). $m_{\gamma}$ regularizes the cross-section in the limit $m_{a}\ll T$. For electrons and positrons in the plasma, it reduces to $m_{\gamma} = eT/3$, see Fig.~\ref{fig:ALP-production-rates}.

In terms of Mandelstam invariants, the squared matrix element takes the form
\begin{equation}
 \frac{g_{i}}{2}g_{\gamma}\overline{|\mathcal{M}|_{i\gamma\to ia}^{2}} = \frac{\pi  Q^{2}_{i}g_{a\gamma\gamma}^2 \alpha_{\text{EM}} \left(t \left(2 m_a^2 m_i^2+2 s m_a^2-m_a^4+4 s m_i^2-2 m_i^4-2 s^2\right)-2 m_a^4
   m_i^2+t^2 \left(2 m_a^2-2 s\right)-t^3\right)}{\left(m_{\gamma }^2-t\right)^2}
\end{equation}
matching Eq. (A.5) from Ref.~\cite{Cadamuro:2010cz}.

\section{ALP decay rates}
\label{app:alp-decay}

The dominant decay mode of the ALP is $a\to \gamma\gamma$. The corresponding width is given by
\begin{equation}
\Gamma_{a\to \gamma\gamma} = \frac{g_{a\gamma\gamma}^{2}m_{a}^{3}}{64\pi}\,,
\label{eq:ALP-dominant-width}
\end{equation}
The sub-leading decay channels are
\begin{equation}
    a\to \gamma + \gamma^{*} \to \gamma + l^{+}l^{-}/\text{hadrons},
    \label{eq:sub-dominant-decays}
\end{equation}
where $\gamma^{*}$ is a virtual photon. Here ``hadrons'' denote a bunch of possible hadronic final states emerging from the coupling of the photon to the hadronic EM current.

The calculation of the width of leptonic decays of the ALPs is straightforward:
\begin{equation}
    \Gamma_{a\to \gamma+l^{+}l^{-}} = \frac{g_{a\gamma\gamma}^{2}\alpha_{\text{EM}}}{96\pi^{2}m_{a}^{3}}\int \limits_{4m_{l}^{2}}^{m_{a}^{2}}d\bar{s} \frac{(m_{a}^{2}-\bar{s})^{3}}{\bar{s}}\sqrt{1-\frac{4m_{l}^{2}}{\bar{s}}}\left(1+\frac{2m_{l}^{2}}{\bar{s}}\right),
    \label{eq:leptonic-width}
\end{equation}
where $\bar{s} \equiv (p_{a}-p_{\gamma})^{2}$. The corresponding branching ratios are highly subdominant, being no more than $1\%$; in addition, electrons and muons do not influence BBN other than by their energy. Because of this, the only impact of these decays is a slight increase in the total ALP decay width, which we neglect for simplicity.

\begin{figure}
    \centering
    \includegraphics[width=0.5\linewidth]{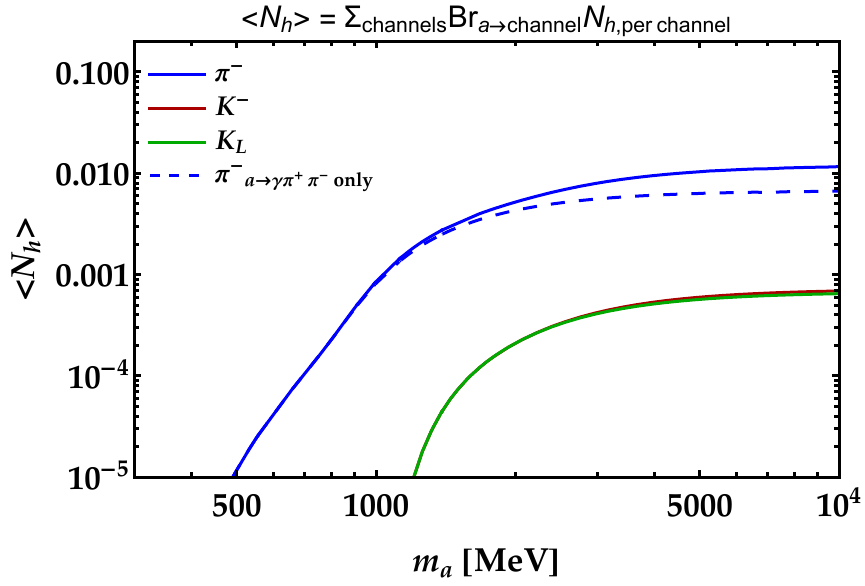}
    \caption{The average number of metastable mesons $h = \pi^{+},K^{+},K_{L}$ per ALP decay, as a function of the ALP mass. The results are obtained by calculating the hadronic decay widths $\Gamma_{a\to \gamma+\text{hadrons}}$, computing the branching ratios $\text{Br}_{a\to \gamma+\text{hadrons}} = \Gamma_{a\to\gamma+\text{hadrons}}/\Gamma_{a,\text{total}}$, and summing over the distinct channels, with the weight $N_{h,\text{channel}}$ being the mean number of mesons $h$ produced per channel. Example: in a decay $a\to \gamma K_{S}K^{*,0}$, $K_{S}$ decays into $\pi^{+}\pi^{-}$ with the probability of $0.692$, while $K^{*,0}$ decays into $\pi^{0}K_{S}\to \pi^{0}\pi^{+}\pi^{-}$ with the probability of $0.692/6$, meaning that $N_{\pi^{-},\gamma K_{S}K^{*,0} }\approx 0.8$. The blue dashed line shows the multiplicity of charged pions if only including the dominant decay mode $a\to \gamma\pi^{+}\pi^{-}$.}
    \label{fig:meson-multiplicities}
\end{figure}

However, the hadronic decays are important, as the metastable mesons $\pi^{\pm},K^{\pm},K_{L}$ that appear in these decays may heavily affect BBN. In particular, they may convert \pn or dissociate nuclei before disappearing. 

In what follows, we discuss the calculation of the branching ratio into these hadronic modes in detail. 

In Fig.~\ref{fig:meson-multiplicities}, we show the summary of our results showing the average number of light mesons per ALP decay.

\subsection{Hadronic rates}
\label{sec:hadronic-rates}
\subsubsection{Induced at tree-level by $a\gamma \gamma$ coupling}
\label{app:alp-decay-tree}
The matrix element of the process is
\begin{align}
    \mathcal{M}_{a\to \gamma+\text{hadrons}} &= \frac{g_{a\gamma\gamma}}{4}\cdot \varepsilon^{\mu\nu\alpha\beta} (p_{\gamma,\mu}\epsilon_{\nu}(p_{\gamma})-p_{\gamma,\nu}\epsilon_{\mu}(p_{\gamma}))\frac{(Q_{\alpha}g_{\beta\kappa} - Q_{\beta}g_{\alpha\kappa})}{Q^{2}}\cdot e J_{\text{EM}}^{h,\kappa}  \nonumber  \\ 
    &=eg_{a\gamma\gamma}\epsilon^{\mu\nu\alpha\beta}p_{\gamma,\mu}\epsilon_{\nu}(p_{\gamma})\frac{Q_{\alpha}}{Q^{2}}J^{h}_{\text{EM},\beta}\,,
    \label{eq:generic-matrix-element}
\end{align}
where $J_{\text{EM},\gamma}^{h}$ is the electromagnetic (EM) hadronic current, $e$ is the EM coupling, $Q \equiv p_{a}-p_{\gamma}$ is the momentum transferred to hadrons, and $\epsilon_{\nu}(p)$ is the photon's polarization vector.

Depending on the scale $Q^{2} = (p_{a}-p_{\gamma})^{2} \equiv \bar{s}$, $J_{\text{EM}}^{h}$ has to be described using different approaches. In the domain $\sqrt{\bar{s}}\gg 2\text{ GeV}$, we are outside the range where intermediate bound-state resonances may influence the results, and the calculation may be done using perturbative QCD, i.e. $J_{\text{EM}}^{h,\delta} = \sum_{q}Q_{q}\bar{q}\gamma^{\delta}q$, where $q$ are quark fields, and $Q_{q}$ is the electric charge. As a result, the ALPs would decay into $\gamma+q+\bar{q}$, with the subsequent showering and hadronization of the $q\bar{q}$ pair. On the other hand, if $\sqrt{\bar{s}}-\sqrt{\bar{s}_{\text{thr}}}\lesssim 4\pi f_{\pi}$, where $\sqrt{s_{\text{thr}}}$ is the threshold energy (being just a sum over the masses of the final hadronic products), perturbative QCD breaks down, and one instead needs to calculate the widths exclusively -- summing over all possible hadronic final states. In the intermediate regime, $4\pi f_{\pi} \lesssim \bar{s}-s_{\text{thr}} \lesssim 2\text{ GeV}$, it is necessary to carefully calculate the rates including all the contributions of the resonances. 

The case of interest for ALPs interacting with photons is the latter one, where resonances are important. The reason is that the matrix element of the process~\eqref{eq:sub-dominant-decays} depends on $\bar{s}$ via the photon propagator, which is maximized at the minimal $Q?$. As a result, hadrons tend to be produced with a smaller invariant mass. This is true even for the ALPs with large masses $m_{a}\gg 1\text{ GeV}$.\footnote{To illustrate this point, we have considered the perturbative width $a\to \gamma q\bar{q}$, and evaluated a fraction of $\bar{s}$ corresponding to the domain $\bar{s}>2\text{ GeV}$. In particular, considering the matrix element as in the perturbative QCD, for the ALP with mass $m_{a} = 10\text{ GeV}$, only $\simeq 5\%$ of the phase space corresponds to $\sqrt{\bar{s}} >2\text{ GeV}$, where perturbative QCD may be applicable.} The intermediate hadronic resonances enhance, however, the distribution at larger $\bar{s}$, so we cannot use ChPT either. 

Fortunately, it is possible to derive the expression for the partial hadronic widths in terms of the experimentally measured quantity
\begin{equation}
    R(s) \equiv \frac{\sigma_{e^{+}e^{-}\to\text{hadrons}}}{\sigma_{e^{+}e^{-}\to \mu^{+}\mu{-}}}\,,
    \label{eq:R-ratio}
\end{equation}
where $s$ is the invariant mass of the colliding $e^{+}e^{-}$ pair. It has the form (see Sec.~\ref{app:derivation-exclusive} for details)
\begin{equation}
    \Gamma_{a\to \gamma+X}=\frac{1}{8\pi^{2}m_{a}^{3}}\int\limits_{4m_{\pi}^{2}}^{m_{a}^{2}}d\bar{s} \  (m_{a}^{2}-\bar{s})^{2}\cdot \sigma_{a+\gamma\to \mu\mu}(\bar{s}) \cdot R_{X}(\bar{s})\,,
\label{eq:exclusive-width}
\end{equation}
with
\begin{equation}
  \sigma_{a+\gamma\to \mu\mu}(\bar{s}) = \frac{\alpha_{\text{EM}}\, g_{a\gamma\gamma}^{2}}{12}\,
  \frac{(m_a^{2}-\bar{s})}{\bar{s}}\,
  \sqrt{1-\frac{4 m_{\mu}^2}{\bar{s}}}\left(1+\frac{2 m_{\mu}^2}{\bar{s}}\right)\,.
\end{equation}

\subsubsection{Derivation of the exclusive widths}
\label{app:derivation-exclusive}

In this subsection, we derive Eq.~\eqref{eq:exclusive-width}. The crucial ingredient is that the hadronic electromagnetic current has non-zero matrix elements between vacuum and one-particle vector meson states $\rho^{0},\omega,\phi, \omega(1420), \dots$ -- the phenomenon known as vector meson dominance~\cite{Sakurai:1960ju,Gell-Mann:1961jim,Kroll:1967it}. Calculating the contributions of these mesons is a non-trivial task, as our knowledge of their properties is limited~\cite{ParticleDataGroup:2024cfk}. Fortunately, it may be possible to calculate the hadronic widths using the experimental data on the scattering $e^{+}e^{-}\to \text{hadrons}$. Using the Hidden Local Symmetry approach of vector meson dominance~\cite{Fujiwara:1984mp}, the data may be expanded onto cross-sections for the partial exclusive hadronic final states~\cite{Ilten:2018crw}: 
\begin{equation}
\text{hadrons} = \pi^{+}\pi^{-}, \ K^{+}K^{-}, K_{L}K_{S}, \ 4\pi, \ \pi^{+}\pi^{-}\pi^{0}, \dots 
\end{equation}
The data is provided in the form of the R-ratios~\eqref{eq:R-ratio}~\cite{ParticleDataGroup:2024cfk} and it may be used to calculate the decay widths of the ALPs $a\to \gamma+\text{hadrons}$ using the ``Dalitz trick''. Namely, there is a relation between the polarization-averaged squared matrix elements of the processes $a \to \gamma+\text{hadrons}$ and $a+\gamma \to \text{hadrons}$:
\begin{equation}
    \overline{|\mathcal{M}_{a\to \gamma+\text{hadrons}}|^{2}} = 2\overline{|\mathcal{M}_{a+\gamma\to \text{hadrons}}|^{2}}\big|_{p_{\gamma}\to -p_{\gamma}}\,.
\end{equation}
The replacement $p_{\gamma}\to -p_{\gamma}$, in particular, switches the invariant mass $s = (p_{a}+p_{\gamma})^{2}$ into $\bar{s} = (p_{a}-p_{\gamma})^{2}$.

Now, let us write the phase space of the final states:
\begin{equation}
    d\Phi_{\gamma,\text{hadrons}} = d\Phi_{\text{hadrons}}\big|_{\sum p_{\text{hadrons}} = p_{a}-p_{\gamma}}\cdot \frac{d^{3}\mathbf{p}_{\gamma}}{(2\pi)^{3}2E_{\gamma}} = d\Phi_{\text{hadrons}}\big|_{\sum p_{\text{hadrons}} = p_{a}-p_{\gamma}}\cdot \frac{(m_{a}^{2}-\bar{s})d\bar{s}}{16\pi^{2}m_{a}^{2}},
\end{equation}
where we used the relation $E_{\gamma}dE_{\gamma} \to (m_{a}^{2}-\bar{s})/(4m_{a}^{2}) d\bar{s}$. Combining the two relations above, for the decay width, we get
\begin{multline}
    \Gamma_{a\to \gamma +\text{hadrons}} = \frac{(2\pi)^{4}}{2m_{a}}\int d\Phi_{\gamma,\text{hadrons}} \cdot \overline{|\mathcal{M}_{a\to \gamma+\text{hadrons}}|^{2}} = \\ = \frac{(2\pi)^{4}}{m_{a}}\int \limits_{4m_{\pi}^{2}}^{m_{a}^{2}}d\bar{s}\ \frac{(m_{a}^{2}-\bar{s})}{16\pi^{2}m_{a}^{2}} \cdot \int d\Phi_{\text{hadrons}}\big|_{\sum p_{\text{hadrons}} = p_{a}-p_{\gamma}}\overline{|\mathcal{M}_{a+\gamma\to \text{hadrons}}|^{2}}\big|_{p_{\gamma}\to -p_{\gamma}}
    \label{eq:ALP-width-temp-0}
\end{multline}
The next step is to express this quantity in terms of the cross-section of the process $a+\gamma\to \text{hadrons}$. We define it with the invariant flux
\begin{equation}
    \sigma_{a+\gamma\to \text{hadrons}}(s) = \frac{(2\pi)^{4}}{2\,\lambda^{\frac{1}{2}}(s,m_{a}^{2},0)}\int d\Phi_{\text{hadrons}}\ \overline{|\mathcal{M}_{a+\gamma\to \text{hadrons}}|^{2}},
    \label{eq:agamma-to-hadrons}
\end{equation}
where $\lambda^{\frac{1}{2}}(s,m_{a}^{2},0)=|s-m_{a}^{2}|$. With this definition (which coincides with the physical cross section for $s\ge m_{a}^{2}$ and provides its analytic continuation for $s\le m_{a}^{2}$), Eq.~\eqref{eq:ALP-width-temp-0} transforms into
\begin{equation}
    \Gamma_{a\to \gamma +\text{hadrons}} = \frac{1}{8\pi^{2}m_{a}^{3}} \int \limits_{4m_{\pi}^{2}}^{m_{a}^{2}} d\bar{s}\ (m_{a}^{2}-\bar{s})^{2}\cdot \sigma_{a+\gamma\to \text{hadrons}}(\bar{s}) \, .
    \label{eq:ALP-width-temp}
\end{equation}
The final step is to relate the hadronic cross-section~\eqref{eq:agamma-to-hadrons} to the R-ratio~\eqref{eq:R-ratio}. The matrix element of the scattering $a+\gamma\to X$ (where $X$ may be any state) is
\begin{equation}
    \mathcal{M}_{a+\gamma\to X} = eg_{a\gamma\gamma}\varepsilon^{\mu\nu\alpha\beta}\epsilon^{*}_{\mu}(p_{\gamma})p_{\gamma,\nu}\frac{Q_{\alpha}}{s}J_{\text{EM},\beta}^{X},
\end{equation}
where $Q = p_{a}+p_{\gamma},Q^{2} \equiv s$, and the current $J_{\text{EM},\beta}^{X}$ was defined in Eq.~\eqref{eq:generic-matrix-element}. The crucial point is that 
\begin{equation}
   F_{\beta\beta'}(Q) \equiv \int d\Phi_{X}J_{\text{EM},\beta}^{X}J^{*}_{\text{EM},\beta'} = -\left(g_{\beta\beta'}-\frac{Q_{\beta}Q_{\beta'}}{s}\right)\int J_{\text{EM},\beta}^{X}J^{X,\beta}_{\text{EM}}d\Phi_{X}\,,
\end{equation}
which follows from the local conservation of the EM current, $Q^{\alpha}J_{\text{EM},\alpha}^{X} = 0$, and the fact that $F_{\beta\beta'}(Q)$ is a covariant function of only one momentum, $Q$. Therefore, we have
\begin{equation}
    \int \overline{|\mathcal{M}_{a+\gamma\to X}|^{2}} d\Phi_{X} =\frac{2\pi\alpha_{\text{EM}}g_{a\gamma\gamma}^{2}(s-m_{a}^{2})^{2}}{s^{2}}\int J_{\beta,\text{EM}}^{X}J^{\beta,X}_{\text{EM}}d\Phi_{X}  \, .
\end{equation}
Now, let us introduce the ratio 
\begin{equation}
   \bar{R}(s) \equiv \frac{\sigma_{a+\gamma\to \text{hadrons}}(s)}{\sigma_{a+\gamma\to \mu\mu}(s)} = \frac{\int J_{\text{EM},\beta}^{h}J^{h,\beta,}_{\text{EM}}d\Phi_{h}}{\int J_{\text{EM},\beta}^{\mu}J^{\mu,\beta}_{\text{EM}}d\Phi_{\mu\mu}}\,,
\end{equation}
where ``$h$'' denotes hadrons. It turns out that $\bar{R}$ matches the experimentally measured $R$-ratio:
\begin{equation}
    R(s) \equiv \frac{\sigma_{e^{+}e^{-}\to h}}{\sigma_{e^{+}e^{-}\to \mu\mu}} = \frac{\int J_{\text{EM},\beta}^{h}J^{h,\beta,}_{\text{EM}}d\Phi_{h}}{\int J_{\text{EM},\beta}^{\mu}J^{\mu,\beta}_{\text{EM}}d\Phi_{\mu\mu}}\,.
\end{equation}
Therefore,
\begin{equation}
\sigma_{a+\gamma\to h}(s) = R(s)\cdot \sigma_{a+\gamma\to \mu\mu}(s)\,.
\end{equation}
The resulting expression for the ALP width $a\to \gamma+X$, Eq.~\eqref{eq:ALP-width-temp}, is
\begin{equation}
    \Gamma_{a\to \gamma+X}=\frac{1}{8\pi^{2}m_{a}^{3}}\int\limits_{4m_{\pi}^{2}}^{m_{a}^{2}}d\bar{s} \  (m_{a}^{2}-\bar{s})^{2}\cdot \sigma_{a+\gamma\to \mu\mu}(\bar{s}) \cdot R_{X}(\bar{s})\,,
    \label{eq:width-via-cross-section}
\end{equation}
where a convenient explicit form for the muon channel (using the invariant-flux definition, valid in the decay region $4m_\mu^2\le \bar{s}\le m_a^2$) is
\begin{equation}
  \sigma_{a+\gamma\to \mu\mu}(\bar{s}) = \frac{\alpha_{\text{EM}}\, g_{a\gamma\gamma}^{2}}{12}\,
  \frac{(m_a^{2}-\bar{s})}{\bar{s}}\,
  \sqrt{1-\frac{4 m_{\mu}^2}{\bar{s}}}\left(1+\frac{2 m_{\mu}^2}{\bar{s}}\right)\,.
  \label{eq:muon-cross-section-continuation}
\end{equation}
We have verified that for the case $X=\mu\mu$, the definitions~\eqref{eq:width-via-cross-section},~\eqref{eq:muon-cross-section-continuation} exactly match the explicit calculation of the width $a\to \gamma\mu\mu$ obtained using Eq.~\eqref{eq:leptonic-width}.

We take the $R$-ratios from~\cite{Ilten:2018crw}. We accounted for the fact that instead of $\sigma_{ee\to \mu\mu}(s)$, the $R$-ratios are normalized by 
\begin{equation}
\bar{\sigma}_{ee\to \mu\mu}(s) = \frac{\sigma_{ee\to \mu\mu}}{\sqrt{1-\frac{4m_{\mu}^{2}}{s}}\left( 1+\frac{2m_{\mu}^{2}}{s}\right)}=\frac{4\pi\alpha_{\text{EM}}^{2}}{3s}\,.
\end{equation}

For the dominant decay $a\to \gamma+\pi^{+}\pi^{-}$, we also utilized the explicit calculation using the form-factor $F_{\pi}(\bar{s})$ obtained using the framework of the extended vector meson dominance fitted by the results of the experimental analysis by BaBar collaboration~\cite{BaBar:2012bdw}, to improve the predictions of the data-driven approach~\eqref{eq:exclusive-width} in the domain $\bar{s} \approx 2m_{\pi}$. Namely, in Eq.~\eqref{eq:generic-matrix-element}, the hadronic current is replaced with
\begin{equation}
    J_{\text{EM},\mu}^{h} = i(p_{\pi^{+},\mu}-p_{\pi^{-},\mu})F_{\pi}(\bar{s})
    \label{eq:pion-form-factor}
\end{equation}
The Lorentz structure of the current follows from the requirement of conservation in the momentum space: $(p_{\pi^{+}}+p_{\pi^{-}})^{\mu}J_{\text{EM},\mu}^{h} = 0$. The definition~\eqref{eq:pion-form-factor} is consistent with Eq.~(24) from Ref.~\cite{BaBar:2012bdw}. Using $F_{\pi}$, we have obtained excellent agreement with the $R$-ratio based calculation~\eqref{eq:width-via-cross-section}.

\section{Thermodynamics of the Universe}
\label{app:thermodynamics}

In this Section, we consider the temperature window $T < T_{\text{split}} = 20\mev$, where the ALPs start to significantly contribute to the energy density of the Universe and also decay, such that the evolution of the ALP and the SM plasma is coupled.

At such temperatures, the plasma may be divided into two components: neutrinos and electromagnetic particles (EM), separated by the dominant interaction handling equilibration. We follow Refs.~\cite{Escudero:2018mvt,EscuderoAbenza:2020cmq,Escudero:2025mvt} to evolve the EM plasma, the neutrino bath, and the scale factor for $T\lesssim\text{MeV}$. The EM temperature is $T_\gamma\equiv T_{\rm EM}$ and we consider that the three active neutrino flavors share a common temperature $T_\nu$; this is because the oscillations equilibrate neutrino flavors already before neutrino decoupling. 

In our case, $N_{\text{eff}}$, the number of effective ultrarelativistic neutrino species is given by:
\begin{equation}
    N_{\text{eff}} \equiv \frac{8}{7}\cdot \left( \frac{11}{4}\right)^{\frac{4}{3}}\frac{\rho_{\nu}}{\rho_{\gamma}}\bigg|_{t\gg t_{\text{ann}},\tau_{a}} =3 \left(\frac{11}{4}\right)^{\frac{4}{3}}\left(\frac{T_{\nu}}{T_\gamma}\right)^{4}\,,
\end{equation}
where $t_{\text{ann}}$ is the time of the electron-positron annihilation. In the standard cosmological scenario, the approach gives $\neff = 3.044$, agreeing well with the unintegrated methods to solve the neutrino Boltzmann equation. $\neff$ is only a well-defined parameter for CMB observations and hence is well defined even for the longest lifetime we considered, $\tau_a = 10^{4}\,{\rm s}$.

\textit{ALP initial conditions -- } For reheating temperatures $\Treh > T_{\rm split}$, the initial ALP number density $n_a(T_{\rm split}) \equiv Y_{a}\cdot s(T_{\text{split}})$ is fixed by the solution of Eq.~\eqref{eq:alp-abundance-equation}, $Y_{a}$ (see Eq.~\eqref{eq:ALP-abundance-final}). In this case, ALPs follow a red-shifted thermal equilibrium distribution depending upon the time of their freeze-out. We define 
\begin{equation}
\mathcal{R}(T)=\frac{\Gamma_a(T)\,n_a^{\rm eq}(T)}{3H(T)\,n_{\rm BE}(T)}\,,
\label{eq:R-criterion-merged}
\end{equation}
where $n_{\rm BE}(T)=\zeta(3)\,T^3/\pi^2$, and where $\Gamma_a(T)$ is the integrated ALP production rate as defined in Eq.~\eqref{eq:ALP-total-production-rate}. If $\mathcal{R}(T_{\rm split})>1$, we consider ALPs to follow an equilibrium Bose-Einstein distribution $f_a^{\rm init}=f_{\rm BE}(E,T_{\rm split})$. Otherwise, we take
\begin{equation}
f_a^{\mathrm{init}}(y)=\frac{n_{a}(T_{\rm split})}{\tilde{n}_{a}}\frac{1}{\exp\left[\frac{\sqrt{m_a^2+\left(m_0\, y/a(\widetilde T)\right)^2}}{\widetilde T}\right]-1}, \quad \tilde{n}_{a} = \int \frac{d^{3}\mathbf{p}}{(2\pi)^{3}}\, \frac{1}{\exp\left[\frac{\sqrt{m_a^2+\left(m_0\, y/a(\widetilde T)\right)^2}}{\widetilde T}\right]-1},
\label{eq:thermodynamics-initial-condition}
\end{equation}
with $\widetilde T$ chosen where $\mathcal{R}(\widetilde T)=1$ (freeze-out) or as $\widetilde T=\Treh$ if production occurs by freeze-in with reheating above $T_{\rm split}$. Finally, $n_{a}(T_{\rm split})$ to ensure the proper normalization on the actual ALP abundance.

We work with comoving momentum variables and define $q\equiv a\,p$ and the auxiliary dimensionless variable $y\equiv q/m_0$ with $m_0$ being an arbitrary constant; then $p=(m_0/a)\,y$. The single-particle energy is $E(p)=\sqrt{p^2+m_a^2}=\sqrt{(q/a)^2+m_a^2}$. Using these conventions, and denoting by $Q_\nu$ the energy exchange rate between neutrinos and electrons and positrons, the full system of equations reads: 
\begin{equation}
\begin{aligned}
\frac{d T_\nu}{dt} &= - H\,T_\nu + \frac{1}{3}\frac{Q_\nu}{d\rho_{\nu}/dT_\nu}\,, \\[2mm]
\frac{d T_\gamma}{dt} &= -\frac{ H\left[4\rho_\gamma + 3\sum_{i=e^\pm,\mu^\pm,\pi^\pm}(\rho_i+p_i)\right] + Q_\nu + Q_a(T_\gamma,a;f_a)}{d\rho_{\rm EM}/dT_\gamma}\,, \\[2mm]
\frac{\partial f_a(t,y)}{\partial t} &= \Gamma_a\!\big(E(p),T_\gamma\big)\,\Big[f_a^{\rm eq}\!\big(E(p),T_\gamma\big)-f_a(t,y)\Big]\,, \qquad p=\frac{m_0}{a}y\,, \\ 
\frac{da}{dt}\frac{1}{a} &= H(t)\,, \qquad
H^2(t) = \frac{8\pi}{3 M_{\rm Pl}^2}\left[\rho_{\rm EM}(T_\gamma)+\rho_\nu(T_\nu)+\rho_a(t)\right]\,, 
\end{aligned}
\label{eq:system-thermodynamics}
\end{equation}
where
\begin{equation}
\rho_{\rm EM}=\rho_\gamma+\rho_{e^\pm}+\rho_{\mu^\pm}+\rho_{\pi^\pm}+\delta\rho_{\rm QED},\qquad
p_{\rm EM}=p_\gamma+p_{e^\pm}+p_{\mu^\pm}+p_{\pi^\pm}+\delta p_{\rm QED}.
\end{equation}
Finite-temperature QED corrections for $\gamma$ and $e^\pm$ up to $\mathcal{O}(e^3)$ are included via $\delta\rho_{\rm QED}$ and $\delta p_{\rm QED}$ following Ref.~\cite{Akita:2020szl}. 

The scale factor is normalized such that $a(T_{\text{split}}) = 1$. Finally, the neutrino exchange $Q_\nu$ is taken from Ref.~\cite{Escudero:2025mvt}.

The ALP energy density and pressure are
\begin{equation}
\rho_a=\int \frac{d^3p}{(2\pi)^3} E\, f_a, \qquad
p_a=\int \frac{d^3p}{(2\pi)^3} \frac{p^2}{3E}\, f_a\,,
\label{eq:moments-merged}
\end{equation}
and the EM$\leftrightarrow$ALP energy exchange is
\begin{equation}
Q_a(T_\gamma,a;f_a)\equiv Q_{\rm EM\to a}=\int \frac{d^3p}{(2\pi)^3} E(p)\,\Gamma_a\!\left(E(p),T_\gamma\right)\left[f_a^{\rm eq}\!\left(E(p),T_\gamma\right)-f_a(p)\right]\,,
\label{eq:Q_flow-merged}
\end{equation}
which is positive when the EM plasma populates ALPs, $f_a<f_a^{\rm eq}$. The momentum-dependent interaction rate sums all the relevant channels at low temperatures:
\begin{equation}
\Gamma_a(E,T_\gamma)=\Gamma_{\gamma\gamma}(E,T_\gamma)+\Gamma_{e\gamma\leftrightarrow ea}(E,T_\gamma),
\label{eq:Gamma_sum-merged}
\end{equation}
with~\cite{Cadamuro:2010cz}
\begin{equation}
\Gamma_{\gamma\gamma}(E,T_\gamma)=\frac{1}{\tau_a}\frac{m_a^2-4m_\gamma^2(T_\gamma)}{m_a^2}\,\frac{m_a}{E}\left[1+\frac{2T_\gamma}{p}\ln\!\frac{1-e^{-(p+E)/(2T_\gamma)}}{1-e^{(p-E)/(2T_\gamma)}}\right]\,,\quad [\text{for $m_a>2m_\gamma(T_\gamma)$}]
\label{eq:Gamma_gg-merged}
\end{equation}
\begin{equation}
\Gamma_{e\gamma\leftrightarrow ea}(E,T_\gamma)=\frac{\alpha_{\rm EM}}{16}\frac{64\pi}{m_a^3\tau_a}\left[4\,n_{e^\pm}(T_\gamma)\right]\ln\!\left(1+\frac{\left[4E\,(m_e+3T_\gamma)\right]^2}{m_\gamma^2(T_\gamma)\left[m_e^2+(m_e+3T_\gamma)^2\right]}\right),
\label{eq:Gamma_egamma-merged}
\end{equation}
where $n_{e^\pm}(T_\gamma)$ is the total $e^\pm$ number density and $m_\gamma(T_\gamma)$ is the plasmon mass given by Eq.~\eqref{eq:mplasmon}.

For the numerical solution, explicitly use the Gauss-Laguerre quadrature method in the dimensionless comoving momentum $y$. With just $\sim 5$ nodes, the method can capture all the relevant effects, and fulfills the continuity equation with a precision of $10^{-5}$ or better at each point in the integration. Our initial conditions for the temperatures are $T_\nu = T_\gamma = 20\mev$ or $T_\nu = T_\gamma = \Treh$ if $\Treh< 20\mev$. For the case of such a low reheating temperature, we start the calculation with $f_a(y) = 0$ (i.e., no ALPs produced before).

In the part of parameter space where ALPs are completely decoupled at $T_{\rm split}$, the full system can be cross-checked with a simplified description. Approximating
\begin{equation}\label{eq:simple}
Q_a \simeq -\frac{\rho_a}{\tau_a \langle\gamma_a\rangle}, \qquad \langle\gamma_a\rangle=\frac{\rho_a}{n_a m_a},
\end{equation}
and evolving $f_a$ only by redshifting, one can solve for $T_\gamma$, $T_\nu$ and $a$ as in Eq.~\eqref{eq:system-thermodynamics} by using $Q_a$ as in Eq.~\eqref{eq:simple} and 
\begin{equation}
\dot\rho_a + 3H\,\rho_a\!\left(1+\dfrac{p_a}{\rho_a}\right) = -\dfrac{\rho_a}{\tau_a\langle \gamma_a\rangle},
\end{equation}
with $p_a/\rho_a$ and $\langle\gamma_a\rangle$ evaluated from the redshifting $f_a^{\rm init}$ in Eq.~\eqref{eq:thermodynamics-initial-condition}. We have explicitly checked that our full results solving for $f_a$ match this description in the domain $m_a > 50\,{\rm MeV}$, providing a good check of the stability of our results.

\section{BBN chain with ALPs}
\label{app:BBN-chain}

Let us introduce the abundance of a hadron $\alpha$: $X_{\alpha} \equiv n_{\alpha}/n_{B}$, where $n_{B}$ is the baryon number density. By definition, for hadrons, $\sum_{hadrons}X_{\alpha} = 1$. In practice, we consider $\alpha = n,d,t,^{3}\text{He},^{4}\text{He},^{7}\text{Li},^{7}\text{Be}$, as these are the only nuclei which have been sizeably produced in the Early Universe. At any moment of time, we may express $X_{p} \equiv 1 - \sum_{H = n,d,\dots}A_{H}X_{H}$. 

The system of equations governing the evolution of the individual abundance $X_{\alpha}$ has the form
\begin{equation}
    \frac{dX_{\alpha}}{dt} = \left(\frac{dX_{\alpha}}{dt}\right)_{\text{weak}}+\left(\frac{dX_{\alpha}}{dt}\right)_{\text{nuclear}} + \left(\frac{dX_{\alpha}}{dt}\right)_{a}
    \label{eq:BBN-chain}
\end{equation}
Here:
\begin{itemize}
    \item $\left(\frac{dX_{\alpha}}{dt}\right)_{\text{weak}}$ is present only for $\alpha = n,p$ and describes the evolution due to the \pn conversion processes mediated by weak interactions. The corresponding processes are
    \begin{equation}
        n+e^{+}\leftrightarrow p + \bar{\nu}_{e}, \quad n + \nu_{e}\leftrightarrow p + e, \quad n \leftrightarrow p + \bar{\nu}_{e}+e^{+},
        \label{eq:weak-processes}
    \end{equation}
    and explicitly, the form of this term is
    \begin{equation}
        \left(\frac{dX_{n}}{dt}\right)_{\text{weak}} = -X_{n}\Gamma^{\rm weak}_{n\to p} + (1-X_{n})\Gamma^{\rm weak}_{p\to n}
    \end{equation}
    The rates $\Gamma^{\rm weak}_{\pn}$ depend on the EM temperature $T$ and the neutrino distribution function $f_{\nu}(p,T)$. As long as neutrinos are in thermal equilibrium with the EM plasma, we have the detailed balance principle: $\Gamma_{n\to p}/\Gamma_{p\to n} \approx \exp\left[(m_{n}-m_{p})/T\right]$, up to the tiny difference between the proton and neutron masses. After neutrino decoupling, the ratio changes -- due to the $e^{+}e^{-}$ annihilation and the effects of the decaying ALPs.
    \item $\left(\frac{dX_{\alpha}}{dt}\right)_{\text{nuclear}}$ is the Standard BBN reaction chain, comprised of interactions between nuclei, nucleons, and photons. The dominant reactions are $2\to 2$ processes, which gives
    \begin{equation}
        \left(\frac{dX_{\alpha}}{dt}\right)_{\text{nuclear}} = \sum_{b,c,d}(-\Gamma_{a+b\to c+d}X_{\alpha}X_{b} + X_{c}X_{d}\cdot\Gamma_{c+d\to \alpha+b}),
        \label{eq:dynamics-nuclear-standard}
    \end{equation}
    where $b,c,d$ are possible interacting particles. The rates $\Gamma_{X\to \alpha+b}$ satisfy the detailed balance principle. In particular, if $b,c,d$ are nuclei/nucleons, we have
    \begin{equation}
        \frac{\Gamma_{X\to \alpha+b}}{\Gamma_{\alpha+b\to X}} = \exp\left[-\frac{Q}{T}\right]\cdot \frac{g_{\alpha}g_{b}}{g_{c}g_{d}}\cdot \left(\frac{m_{a}m_{b}}{m_{c}m_{d}}\right)^{\frac{3}{2}} \cdot \frac{S_{cd}}{S_{ab}}
    \end{equation}
    Here, $m_{y}$ is the mass of the particle $y$, $g_{y}$ is the number of its internal degrees of freedom (helicities), while $S_{y_{i}y_{j}}$ is the combinatoric coefficient being $2$ if $i=j$ and 1 otherwise.
    \item $\left(\frac{dX_{\alpha}}{dt}\right)_{a}$ is the evolution due to the presence of metastable mesons $\pi^{\pm},K^{\pm},K_{L}$, that appear among the ALP decay products. Explicitly,
    \begin{equation}
        \left(\frac{dX_{\alpha}}{dt}\right)_{a} = \sum_{h}n_{h}\langle \sigma v\rangle_{\alpha+h\to \{y\}},
    \end{equation}
    with $n_{h}$ being the instant number density of the mesons, and $\{y\}$ being some final state. There is no backward reaction because all the mesons instantly disappear from the plasma.
\end{itemize}

The dynamics of the scale factor and the time-temperature relation are obtained using the system~\eqref{eq:system-thermodynamics}.

We discuss the ingredients needed to derive the summands in~\eqref{eq:BBN-chain} below, in Sec.~\ref{sec:BBN-chain-details}. We have validated the resulting framework by the following tests: reproducing SBBN; reproducing the parameter space shown in Fig.~5 of Ref.~\cite{Pospelov:2010cw}; and obtaining the asymptotic constraint on the ALP lifetimes coming from the meson-driven \pn conversion, $\tau_{a}\lesssim 0.02\s$, which generically occurs in studies of hadronically decaying relics~\cite{Kawasaki:2004qu,Fradette:2018hhl,Boyarsky:2020dzc}.

\subsection{Derivation of the BBN chain terms}
\label{sec:BBN-chain-details}

\textbf{Weak-driven processes.} Under our approximation of the neutrino population, $f_{\nu}(p,T) \equiv f_{\text{FD}}(T_{\nu_{e}}(T),p)$, where the evolution $T_{\nu_{e}}(T)$ is given by the system~\eqref{eq:system-thermodynamics}. Having this, we first calculate the bare \pn rates of the processes~\eqref{eq:weak-processes}. The overall constant in the bare rates is normalized by the neutron lifetime. Then, we multiply them by the temperature-dependent factor incorporating various corrections coming from QED, finite nucleon mass, and others. We take the latter from the PRIMAT repository~\cite{Pitrou:2018cgg}.

\textbf{Standard BBN chain}. The rates entering the nuclear dynamics~\eqref{eq:dynamics-nuclear-standard}, have explicit form $\Gamma_{a+b\to c+d} = n_{B}\cdot \langle\sigma v\rangle_{a+b\to c+d}$. We take the averaged cross-sections $\langle\sigma v\rangle_{a+b\to c+d}$ from the PRIMAT repository. The baryon number density is expressed as
\begin{equation}
    n_{B} \equiv \eta_{B}(T)\cdot n_{\gamma}(T),
\end{equation}
where $n_{\gamma} = 2\zeta(3)/\pi^{2}T^{3}$ is the number density of photons, and 
\begin{equation}
    \eta_{B} = \eta_{B,\text{rec}}\cdot \left( \frac{a(T_{\text{rec}})T_{\text{rec}}}{a(T)T}\right)^{3}
\end{equation}
is the temperature-dependent baryon-to-photon ratio, fixed by the value $\eta_{B,\text{rec}} = 6.109\cdot 10^{-10}$ at the recombination epoch, which is extracted from the CMB measurements.

\textbf{Meson-driven processes.} For incorporating the meson-driven processes, we mainly follow Refs.~\cite{Reno:1987qw,Pospelov:2010cw,Akita:2024nam,Akita:2024ork}. 

For the \pn conversion processes, we consider
\begin{align}
    \pi^{-}+p &\to n+\gamma/\pi^{0}, \quad \pi^{+}+n\to p+\pi^{0}, \label{eq:pn-conversion-pions} \\ K^{-}+p&\to n+m\pi, \quad K^{-}+n(p)\to p(n)+m\pi, \quad K_{L}+n(p)\to p(n)+m\pi,
    \label{eq:pn-conversion-kaons}
\end{align}
where $m \pi$ is the final state comprised of $m$. The kaon-driven reactions are mediated by the intermediate resonant states $\Sigma\pi /\Lambda \pi$, with the final states comprised of $m$ pions.  Interestingly, the process with the final photon is present for the incoming $\pi^{-}$ but not $\pi^{+}$. It follows from the fact that at close to threshold, $\pi^{-}p$ scattering occurs via an intermediate pionic Hydrogen, which has comparable radiative and pionic decay modes~\cite{Panofsky:1950he}. On the other hand, for the $\pi^{+}n$ scattering, there is no such intermediate state.

For the nuclear dissociation processes, we use
\begin{align}
    \pi^{-}+^{4}\text{He} &\to t + n, \quad \pi^{-}+^{4}\text{He}\to d + 2n, \quad \pi^{-}+^{4}\text{He}\to p+3n, \\ K^{-}+^{4}\text{He}&\to ^{3}\text{He}+m\pi, \ K^{-}+^{4}\text{He}\to \text{t}+n+m\pi, \\ K^{-}+^{4}\text{He}&\to \text{d}+2n+m\pi, \ K^{-}+^{4}\text{He}\to p+3n+m\pi
\end{align}
The rates of the processes above with oppositely charged particles are enhanced because of the Coulomb attraction, parametrized by the Sommerfeld factor
\begin{equation}
    F_{C}(T) = \frac{x}{1-\exp[-x]}, \quad x= \frac{2\pi\alpha_{\text{EM}}Z}{v}, 
\end{equation}
where $Z$ is the electric charge of the we approximate the velocity with $v = \sqrt{2T/\mu}$ and $\mu$ being the reduced mass. The reactions of the type $\pi^{+}+^{4}\text{He}\to X$ are, vice versa, suppressed by the Coulomb repulsion, and for this reason, following Ref.~\cite{Pospelov:2010cw}, we do not include them. Finally, we do not include the meson-mediated dissociation processes of rare nuclei $d, t, ^{3}\text{He}, ^{7}\text{Li}, ^{7}\text{Be}$: compared to the $^{4}\text{He}$ dissociation, the rates of such processes are highly suppressed by the nucleis' densities and therefore are sub-dominant. 

The meson-driven rates are defined by
\begin{equation}
    \Gamma_{a+h\to y} = n_{h} \cdot \langle \sigma v\rangle_{h+a\to y},
\end{equation}
where $n_{h}$ is the meson's instant number density and $\langle \sigma v\rangle_{a+h\to y}$ is the cross-section of the process $a+h \to y$ with some final state $y$ averaged over $h$'s energies. We discuss the evolution of $n_{h}$ and the calculation of $\langle \sigma v\rangle_{h+a\to y}$ below.

\subsection{Meson-driven rates}

\textbf{Meson population evolution.} Let us discuss the evolution of the mesons' population. It follows from the combination of the injection by decaying ALPs, kinetic energy loss in elastic EM scattering, decays, self-annihilations of $h$ and anti-$h$, and scattering off nucleons (including not only conversions but also the reactions that do not change the nucleon type). This cascade couples the dynamics of various mesons~\cite{Akita:2024nam,Akita:2024ork}. Namely, reactions with $K^{\pm},K_{L}$ produce pions and muons; on the other hand, the $K$ dynamics is sensitive to the baryon-to-photon ratio, which is affected by the pions. 

Moreover, if mesons have a chance to decay, they may inject high-energy neutrinos, which non-trivially influence the neutrino thermalization. This makes the resulting evolution of the plasma bath very complicated. In particular, it may lead to the necessity of considering the dynamics of the nucleons and the SM plasma simultaneously, as the nucleons control the dynamics of mesons, which, in turn, determines the energy distribution between the EM and neutrino plasma components.

In our case, however, we have 
\begin{equation}
\langle N_{K}\rangle\ll \langle N_{\pi^{\pm}}\rangle \lesssim 10^{-3}
\label{eq:meson-yields-smallness}
\end{equation}
recall Fig.~\ref{fig:meson-multiplicities}. This regime allows us to perform three simplifications:
\begin{enumerate}
\item We can neglect the self-annihilations. Indeed, their rate is proportional to the yield of the anti-meson, $\langle N_{h}\rangle$. The self-annihilation dominates the meson dynamics if $\langle N_{h}\rangle\gtrsim 0.1$; hence, they are negligible at MeV temperatures. 
\item We may factorize the evolution of different mesons. It is reasonable because reactions with kaons would only produce a tiny amount of pions (due to $\langle N_{K}\rangle\ll \langle N_{\pi^{\pm}}$), and hence the kaon source term in the equation of the evolution of pions may be neglected. The only remaining effect coupling the mesons' dynamics is the sink term handling the interaction with nucleons -- it is proportional to the neutron abundance $X_{n}$, which is $n_{\pi,K}$-dependent. However, $X_{n}$ stays around $0.5$ independently of the meson palette~\cite{Boyarsky:2020dzc,Akita:2024ork}; further, we approximate $X_{n} = 0.5$ in this term.
\item We may neglect the injection of non-thermal neutrinos: due to~\eqref{eq:meson-yields-smallness}, they would carry the energy density $\ll 1\%$, which effectively makes their impact on \neff invisible.
\end{enumerate}

Explicitly, the resulting instant number density of mesons is given by
\begin{equation}
    n_{h} \approx \frac{n_{a}}{\tau_{a}}\cdot \langle N_{h}\rangle \cdot \frac{1}{\tau_{h}^{-1}+\sum_{N = n,p}n_{N}\langle \sigma v\rangle_{h+N\to y}}
    \label{eq:meson-instant-number-density}
\end{equation}
Here, $\tau_{h}$ is the meson's lifetime, while $\langle \sigma v\rangle_{h+N\to N(N') + y}$ is the total meson-driven cross-section of interactions with nucleons, including \pn conversion and quasi-elastic processes $h+N\to N+\dots$ where the meson $h$ disappears (with producing lighter particles $\dots = \pi^{0},\gamma$, etc.). Finally, $n_{N}$ is the number density of the nucleon $N$.

The second summand in the denominator of Eq.~\eqref{eq:meson-instant-number-density}, coming from the interaction with nucleons, may dominate over decays at temperatures $T\gg 1\mev$. However, it scales as $n_{B}\propto T^{3}$, and becomes negligible compared to the decay rate already at $T \simeq 0.5\mev$. For the same reason, we do not include a similar sink term due to the interactions with nuclei: it only becomes non-negligible compared to the nucleon-driven terms at $T\lesssim 80\text{ keV}$, where all hadronic interactions are much slower than the decay rate.

\textbf{Meson population evolution.} To know the averaged cross-section of meson interactions, we need to know the energy distribution of mesons throughout their evolution since their injection.

Immediately after being injected, they have the energy distribution specified by the ALP mass. The elastic EM interactions, mainly through Coulomb scatterings off electrons and the inverse Compton process, lead to the loss of this energy. Depending on whether the energy loss rate is faster than the interaction with hadrons, the mesons either end up having thermal kinetic energy distribution (i.e., effectively at rest at temperatures $T\lesssim 1\mev$) or while being incompletely stopped. Further, we will consider two temperature ranges, $T \gtrsim 40\text{ keV}$ and $T<40\text{ keV}$, defined by the strength of kinetic energy loss of the mesons~\cite{Pospelov:2010cw}. 

At temperatures $T \gtrsim 40\text{ keV}$, the charged mesons instantly lose their kinetic energy. Since all the hadronic interaction processes we consider above are thresholdless, we may approximate their cross-section by the ``thermal cross-section'' for the stopped mesons $\langle \sigma v\rangle^{\text{therm}}_{a+h\to y}$:
\begin{equation}
\langle \sigma v\rangle_{a+h\to y} \approx \langle \sigma v\rangle^{\text{therm}}_{a+h\to y} = (\sigma v)_{\text{thr}} \cdot F_{C}(T), \quad T \gtrsim 40\text{ keV}
\label{eq:cross-section-approximation-threshold}
\end{equation}
where $(\sigma v)_{\text{thr}}$ is the bare hadronic cross-section for $E_{h} = m_{h}$ without the Coulomb factor. $(\sigma v)_{\text{thr}}$ can be extracted from pionic atom lifetime data and mesonic capture by Helium~\cite{Pospelov:2010cw}. For the collision of two particles where one is chargeless, $F_{C}$ has to be replaced with 1.

Once the temperature lowers, the EM scattering rates decrease. At temperatures $T<40\text{ keV}$, they become comparable with the meson decay rate. As a result, we must consider the effects of finite kinetic energy, and in particular, the enhanced lifetime of the mesons and increased interaction cross-section. All the processes except for the pion-driven $\pn$ conversion and $^{4}\text{He}$ dissociation~\eqref{eq:pn-conversion-pions} are very far from threshold, which means that for them, the approximation~\eqref{eq:cross-section-approximation-threshold} is still reasonable with $\mathcal{O}(1)$ accuracy. For the latter two reactions, the approximation breaks down. The main reason (apart from the enlarged phase space) is the intermediate $\Delta$ resonance in the scattering process 
\begin{equation}
\pi^{-}+p\to n+\pi^{0}, \quad \pi^{+}+n\to p+\pi^{0}, \quad \pi^{-}+^{4}\text{He}\to X
\end{equation}
The cross-sections of these reactions are maximized at the kinetic energy $K_{\pi} \equiv E_{\pi}-m_{\pi} \approx 180\mev$, such that the scale of the transferred momentum is close to $m_{\Delta}$. It leads to a necessity for studying the dynamics of the pion energy loss.

\begin{figure}[h!]
    \centering
    \includegraphics[width=0.5\linewidth]{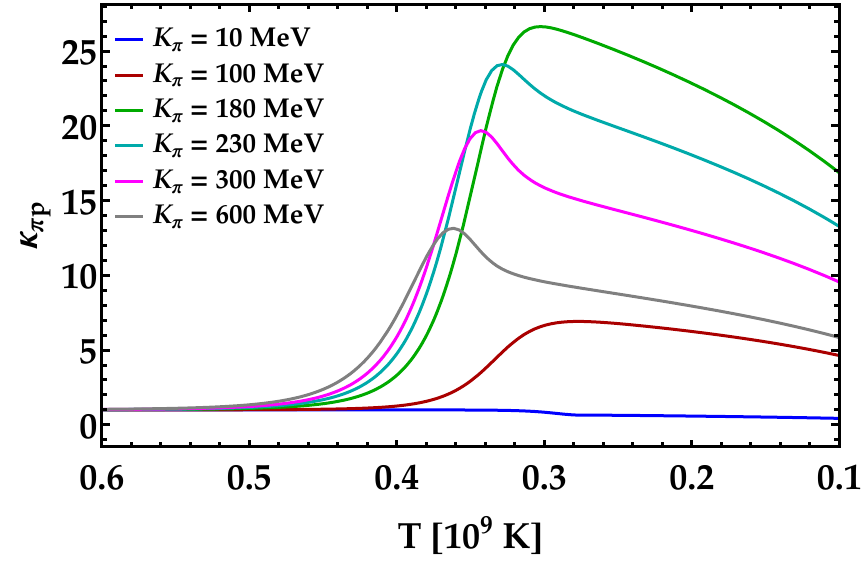}~ \includegraphics[width=0.5\linewidth]{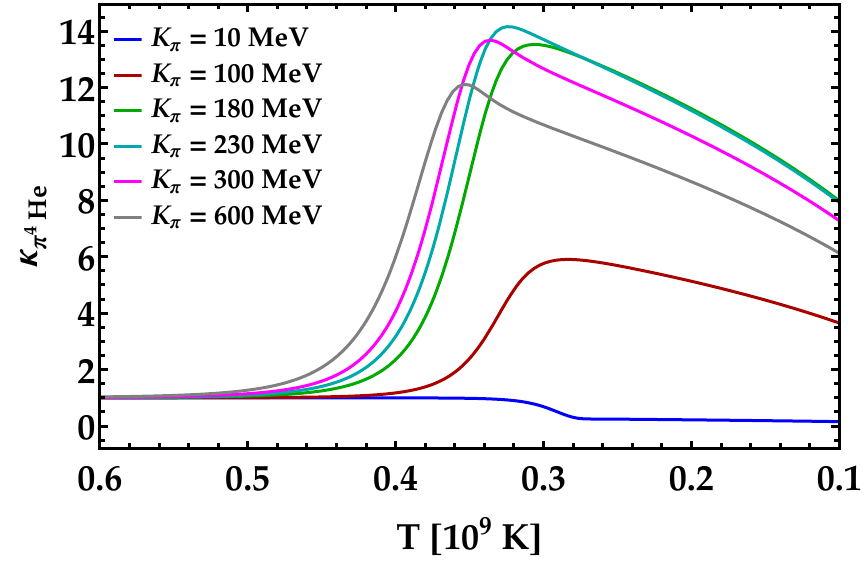}
    \caption{The enhancement of the probability to interact with hadrons for the charged pions injected at various temperatures $T$ with different kinetic energies $K_{\pi}$.}
    \label{fig:kappa-factors}
\end{figure}

To calculate it, we follow Ref.~\cite{Pospelov:2010cw}. First, let us write 
\begin{equation}
    n_{\pi}\cdot \langle \sigma v\rangle_{\pi A \to X} \equiv n_{a}\frac{\tau_{\pi}}{\tau_{a}}\langle N_{\pi}\rangle \cdot \langle \gamma_{\pi}\sigma v\rangle_{\pi} \equiv n_{a}\cdot \langle N_{\pi}\rangle \cdot \tau_{a}\langle \sigma v\rangle^{\text{therm}}_{\pi A\to X}\cdot \kappa(K_{\pi},T),
\end{equation}
where the parameter $\kappa$ encapsulates the effect of incomplete stopping of pions:
\begin{equation}
    \kappa_{\pi A}(K_{\pi},T) \equiv \frac{\int \limits_{0}^{\infty} dt (\sigma v)_{\pi A\to X}(\mathcal{K}_{\pi}(t))\cdot \exp\left[-\int \limits_{0}^{t}dt' \frac{1}{\tau_{\pi}\gamma(\mathcal{K}_{\pi}(t'))}\right]}{\tau_{a}\langle \sigma v\rangle^{\text{therm}}_{\pi A\to X}}
    \label{eq:kappa}
\end{equation}
Here, $(\sigma v)_{\pi A\to X}(\mathcal{K}_{\pi})$ is the energy-dependent interaction cross-section. $\kappa$ is driven by the thermalization of pions (the evolution of their kinetic energy $\mathcal{K}_{\pi}(t)$) in Coulomb and inverse Compton scattering off background electrons and photons:
\begin{equation}
    t(\mathcal{K}_{\pi}) = \int \limits^{K_{\pi}}_{\mathcal{K}_{\pi}}\frac{dK_{\pi}'}{|dE'/dt|}, \quad \frac{dE'}{dt} = \left( \frac{dE'}{dt}\right)_{\text{Coulomb}}+\left( \frac{dE'}{dt}\right)_{\text{Compton}}
\end{equation}
The limiting values for $\kappa$ are
\begin{equation}
    \kappa_{\pi A}(K_{\pi},T) \to \begin{cases} 1, \qquad T \gg 40\text{ keV}, \\  \gamma(K_{\pi})\frac{(\sigma v)_{\pi A\to X}(K_{\pi})}{\langle \sigma v\rangle^{\text{therm}}_{\pi A\to X}}, \qquad T\ll 10\text{ keV}  \end{cases}
\end{equation}
If $T\gg 40\text{ keV}$, the thermalization is complete, so $\kappa \to 1$. For $T\ll 10\text{ keV}$, the stopping turns off, and pions scatter/decay with the kinetic energy they were produced.

\begin{figure}[h!]
    \centering
    \includegraphics[width=0.5\linewidth]{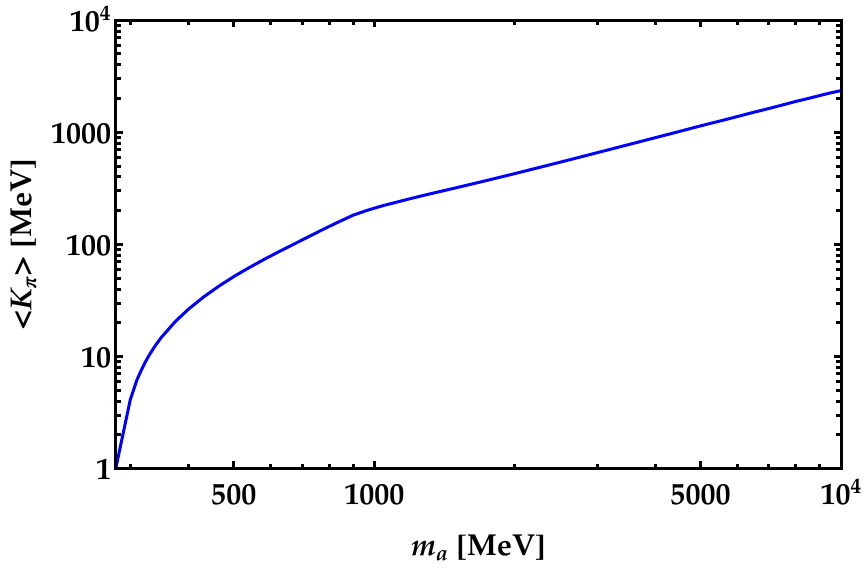}
    \caption{The mean kinetic energy of the pion $\langle K_{\pi}\rangle$ as calculated by the dominant pion production mode $a\to \pi^{+}\pi^{-}\gamma$. See Sec.~\ref{app:alp-decay} for details.}
    \label{fig:pion-mean-energy}
\end{figure}

In the domain $10\text{ keV}\lesssim T \lesssim 40\text{ keV}$, the thermalization is present but is incomplete, which, in light of the $\Delta$-driven enhancement, introduces a non-trivial dynamics in $\kappa$. As shown in Ref.~\cite{Pospelov:2010cw}, depending on the initial kinetic energy $K_{\pi}$, the $\kappa$ factor may be as large as a factor of 30.

To calculate $\kappa$, we approximate $K_{\pi}$ by its average value in ALP decays $\langle K_{\pi}\rangle(m_{a})$. To this extent, we used the differential width of the dominant hadronic decay process $a\to \pi^{+}\pi^{-}\gamma$, which we calculate in terms of the effective $\gamma\pi\pi$ form-factor (recall Eq.~\eqref{eq:pion-form-factor}). The behavior of this average energy as a function of the ALP mass is shown in Fig.~\ref{fig:pion-mean-energy}. Next, we have utilized Eq.~\eqref{eq:kappa}, with the pion-energy-dependent cross-sections taken from Appendix A.2 of Ref.~\cite{Pospelov:2010cw}. We have validated the fit by reproducing Fig.~2 of this work, see Fig.~\ref{fig:kappa-factors}.

For the process $\pi^{+}+n\to p+\pi^{0}$, the authors did not consider the enhancement factor, motivated by the fact that at times when $\pi^{+}$ may not be stopped, all the neutrons become bound inside light nuclei. However, to have a consistent picture of nuclear dynamics in the presence of mesons, it may be useful to include the corresponding enhancement as well. Given the isospin symmetry, we take the same cross-section as for the $\pi^{-}+p\to n+\pi^{0}$, but rescaling it with the appropriate factor for tiny kinetic energies, to account for the different release energy of the $\pi^{-}p$ and $\pi^{+}n$ processes at threshold and maintain equal cross-sections in the large $K_{\pi}$ limit.

\section{Extra Results}
\label{app:some-results}

\begin{figure}[h!]
    \centering
    \includegraphics[width=0.5\linewidth]{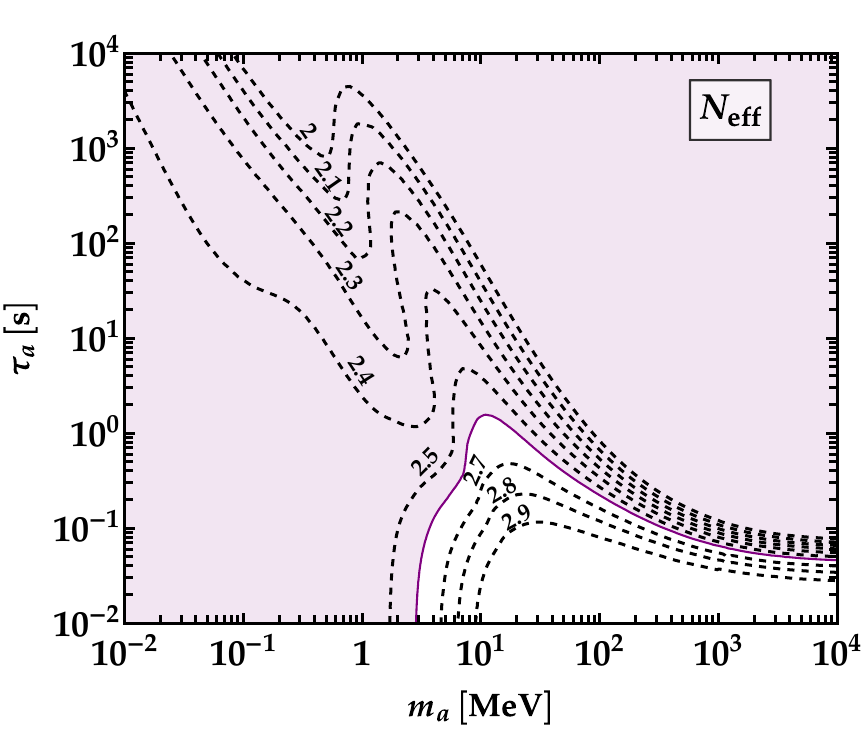}~\includegraphics[width=0.5\linewidth]{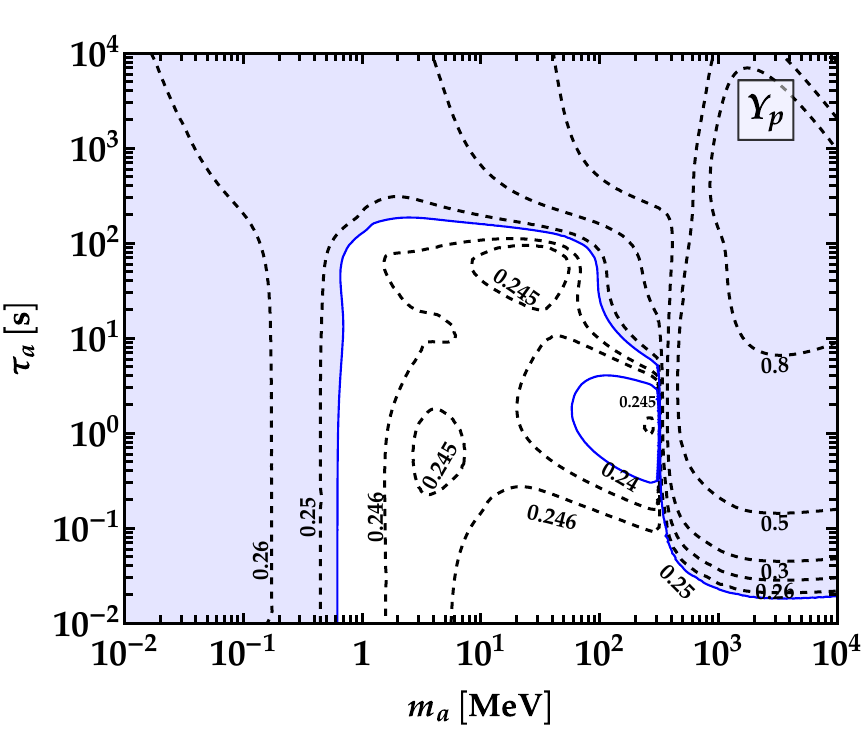}\\ \includegraphics[width=0.5\linewidth]{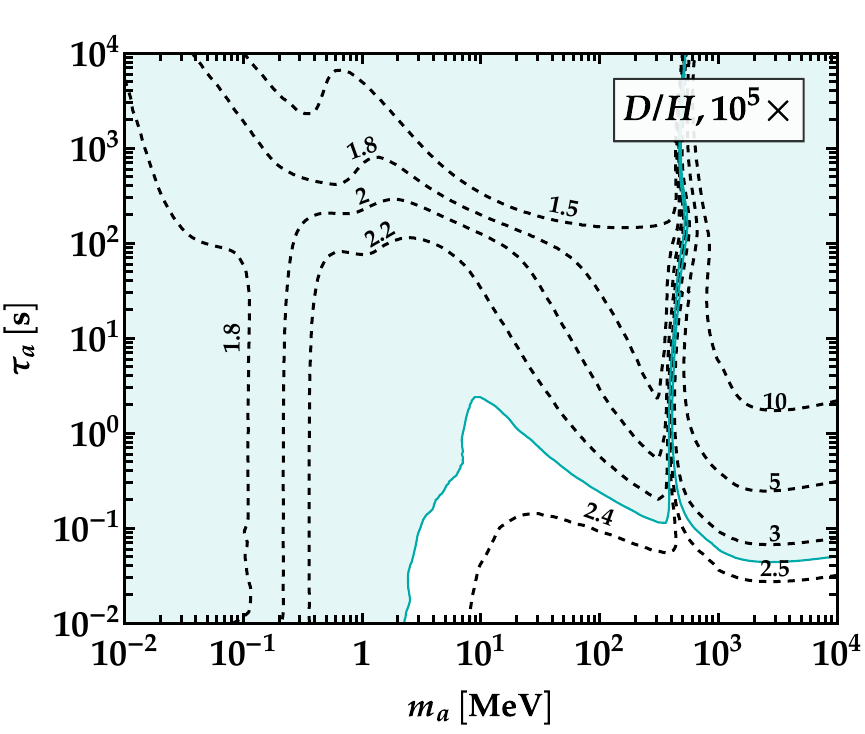}
    \caption{The ALP parameter space in terms of the ALP mass $m_{a}$ and lifetime $\tau_{a}$ mapped onto iso-contours of constant values of \neff, the helium-4 abundance \yp, and the D-to-H ratio \dH. Here, we consider $\Treh = 10^{10}\gev$. The colored domains correspond to $2\sigma$-excluded regions, as discussed in Sec.~\ref{sec:methodology}. Note that we ran our calculations for fixed $\Omega_bh^2 = 0.02242$ and this gives $N_{\rm eff}^{\rm SM} = 3.044$ and with our nuclear reaction rates: $10^5\,{\rm D/H}|_{\rm P}^{\rm SM} = 2.44$ and $Y_{\rm P}^{\rm SM} = 0.247$. {Finally, the dashed black lines in these figures denote the iso-contours of fixed \neff, \yp, and \dH, correspondingly.}}
    \label{fig:iso-contours}
\end{figure}

Here, we show some further details and results from our calculations. Fig.~\ref{fig:iso-contours} shows iso-contours of \neff, \dH, and \yp for the case of $T_{\rm reh} = 10^{10}\,{\rm GeV}$ together with our $2\sigma$ limits. 

\textbf{Number of effective relativistic neutrino species.} In the presence of electromagnetically decaying ALPs, \neff is generically smaller than its SBBN value $N_{\text{eff,SBBN}}\simeq 3.044$. However, the lifetime span of sizable $\Delta \neff = \neff - N_{\text{eff,SBBN}}<0$ significantly varies with the ALP mass. Very low-mass ALPs, $m_{a}\lesssim 1\mev$, are in thermal equilibrium with the SM bath at MeV temperatures and disappear after neutrinos decouple. This way, they may affect \neff even for the lifetimes $\tau_{a}\lesssim 0.01\s$, where a thermal relic that decoupled earlier would not have affected the observables. This results in a non-zero correction $\Delta\neff$ in this lifetime domain. 

For larger masses, ALPs typically behave as thermal relics and may only modify \neff if at least a small fraction of them decay at $T\lesssim 2\mev$, during the neutrino decoupling. In practice, it means that only the lifetimes $\tau_{a}\gtrsim 0.04\s$ may sizably change \neff.

\textbf{Helium abundance. }The behavior of \yp changes significantly as a function of the ALP mass. In particular, the shift with respect to the standard model value $Y_{\rm P}^{\rm SM} = 0.247 $ may be positive or negative. For very small masses $m_{a}\lesssim 1\mev$, $\Delta \yp$ is positive -- mainly due to accelerating the expansion of the Universe by the ALPs that are in (partial) equilibrium with the SM plasma. In the mass range $1\mev\lesssim m_{a}\lesssim 2m_{\pi}$, we enter the opposite regime where the ALPs behave rather as thermal relics. Their combined effect on the expansion of the Universe and the \pn conversion results in a smaller $n/p$ ratio and hence a smaller \yp. These effects are most notable for $\tau_a \sim 0.1-10\,{\rm s}$ for $m_a \sim 20-200\,{\rm MeV}$ and can led to $Y_{\rm P}$ as low as $Y_{\rm P} = 0.24$. In the domain of larger masses, $m_{a}\gtrsim 2m_{\pi}$, the dominant effect is the meson-driven \pn conversion, which increases the $n/p$ ratio and hence leads to a larger \yp.

\textbf{Deuterium abundance. } The \dH pattern is somewhat similar to that of \yp but with an important difference: the interplay of effects seen for $Y_{\rm P}$ for $0.1\,{\rm s}\lesssim\tau_a \lesssim 100\,{\rm s}$ and $1\,{\rm MeV}\lesssim m_a \lesssim 250\,{\rm MeV}$ is not seen. The reason is that $Y_{\rm P}$ is both sensitive to the expansion rate of the Universe at $T_\gamma \simeq 0.7\,{\rm MeV}$ and at $T_\gamma \simeq 0.075\,{\rm MeV}$ but is also affected by the proton-to-neutron conversion rates that are affected by the ALP decays. On the other hand, the deuterium abundance in this part of the parameter space is primarily sensitive to the expansion rate of the Universe at $T_\gamma \simeq 0.075\,{\rm MeV}$ and hence we see that the exclusion region ends up following quite closely the one for $N_{\rm eff}$.

\section{Comparison with the previous works}
\label{app:previous-works-comparison}

In this section, we compare the results of our analysis with the previous studies~\cite{Cadamuro:2011fd,Depta:2020wmr}. Apart from different criteria of constraints tied to different observational status of primordial nuclear abundances and Cosmic Microwave Background, another difference is attributed to distinct physics input, which we comment on below.

\begin{figure}[h!]
    \centering
    \includegraphics[width=0.5\linewidth]{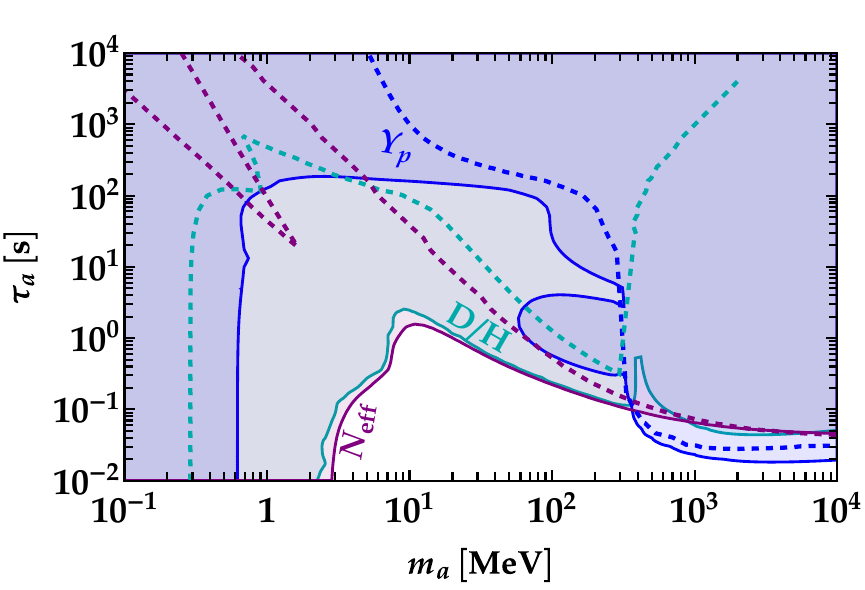}
    \caption{Comparison of the results of our work with Fig. 2 of Ref.~\cite{Cadamuro:2011fd}. The colored domains denote our calculations: the blue one shows the bounds on primordial helium abundance ($Y_{\rm P})$, the cyan domain is the bound on the deuterium abundance (${\rm D/H}|_{\rm P}$), while the purple domain depicts \neff bounds. The dashed lines are corresponding bounds of Ref.~\cite{Cadamuro:2011fd}, with everything excluded on top.}
    \label{fig:comparison-cadamuro}
\end{figure}
Let us start with Ref.~\cite{Cadamuro:2011fd}. The authors have performed the analysis assuming effectively infinite reheating temperature \Treh, and included the meson-driven nucleon conversion and nuclei dissociation. The comparison is shown in Fig.~\ref{fig:comparison-cadamuro} in terms of the final limits in the $\tau_a$ vs $m_a$ plane. 

In the domain $m_{a}\gtrsim 2m_{\pi}$, where the dynamics of the helium abundance is completely dominated by the meson-driven processes, the helium abundance constraints as derived in both studies are very similar. For lower masses, sizable deviations appear, with the bound from Ref.~\cite{Cadamuro:2011fd} appearing only at very large lifetimes $\tau_{a}\gtrsim 100\s$. The main reason is that Ref.~\cite{Cadamuro:2010cz} imposed the conservative bound on the helium abundance from above only. {Furthermore, there is a bump at $\tau_a \sim 1\,{\rm s}$ which is now constrained and we attribute this to the very precise $Y_{\rm P}$ measurement we are using, see Eq.~\eqref{eq:BBN_data}.} For masses $m_{a}\lesssim 2m_{\pi}$, the combined effect of the EM decays and the influence on the expansion of the Universe either leads to a decrease in the helium abundance for the lifetimes $\tau_{a}\lesssim 100\s$.

As for the ${\rm D/H}|_{\rm P}$ limits, the discrepancy in the domain $m_{a}\gtrsim 2m_{\pi}$ is related to the way Ref.~\cite{Cadamuro:2011fd} formulated their ${\rm D/H}|_{\rm P}$ bound -- as the conservative bound from below. The meson-driven effects tend to increase ${\rm D/H}|_{\rm P}$, and hence the ${\rm D/H}|_{\rm P}$ bounds from Ref.~\cite{Cadamuro:2011fd}. In contrast, we use the 95\% CL domain of the values of ${\rm D/H}|_{\rm P}$ allowed by observations, which includes the upper boundary.

For the \neff bounds, the major, qualitative discrepancy lives in the domain $m_{a}\lesssim 10\mev$, where the constraints of Ref.~\cite{Cadamuro:2011fd} quickly weaken at small lifetimes, while our constraints continuously extend toward smaller lifetimes. The reason is attributed to the boundary on $N_{\text{eff}}$ the authors defined from observations: $N_{\text{eff}}>2.11$. From their analysis, in the domain of small ALP masses $m_{a}\lesssim 10\mev$, such small values are only possible if increasing the lifetime.

Finally, we can also compare our results in Fig.~\ref{fig:iso-contours} for the various observables to the results in Figs. 3 and 4 of~\cite{Cadamuro:2012rm} finding very good agreement in the overall shapes of the various iso-contours.

\begin{figure}[h!]
    \centering
    \includegraphics[width=0.5\linewidth]{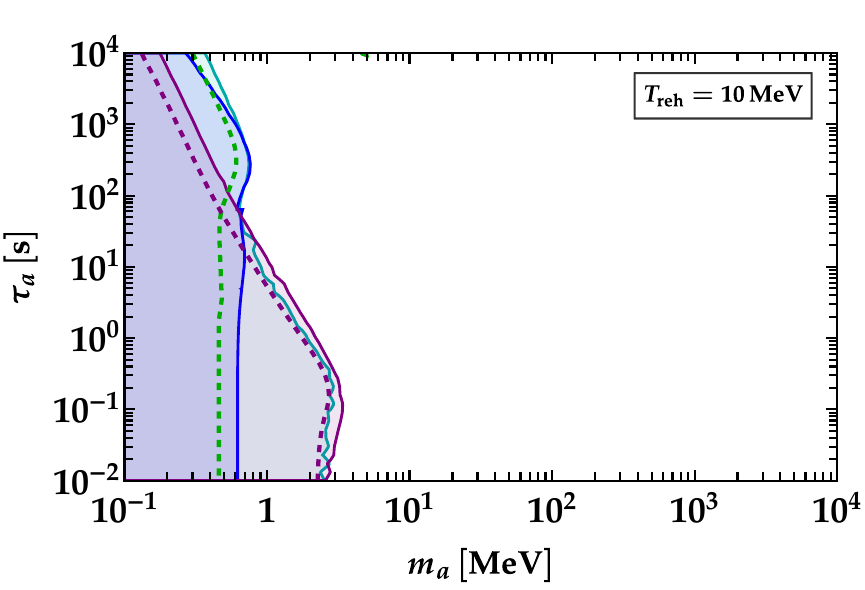}~\includegraphics[width=0.5\linewidth]{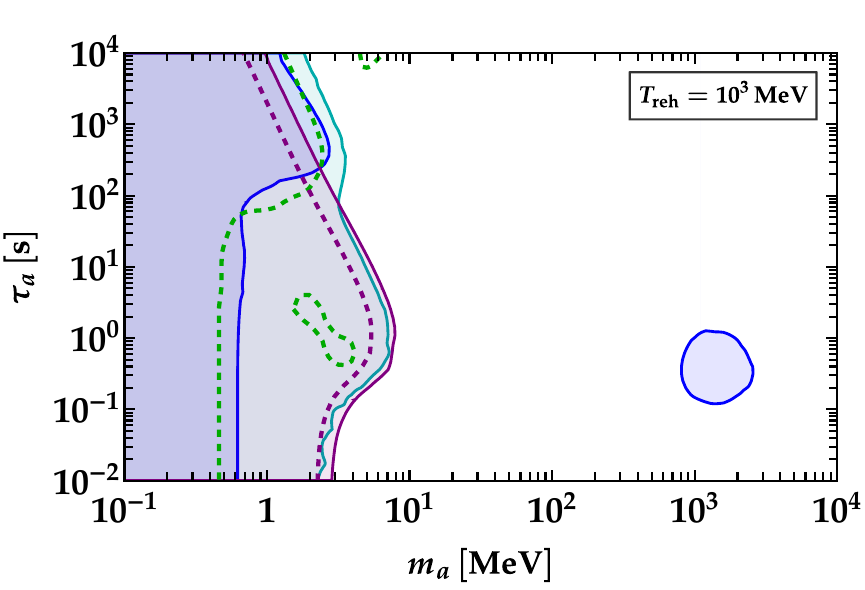} \\   \includegraphics[width=0.5\linewidth]{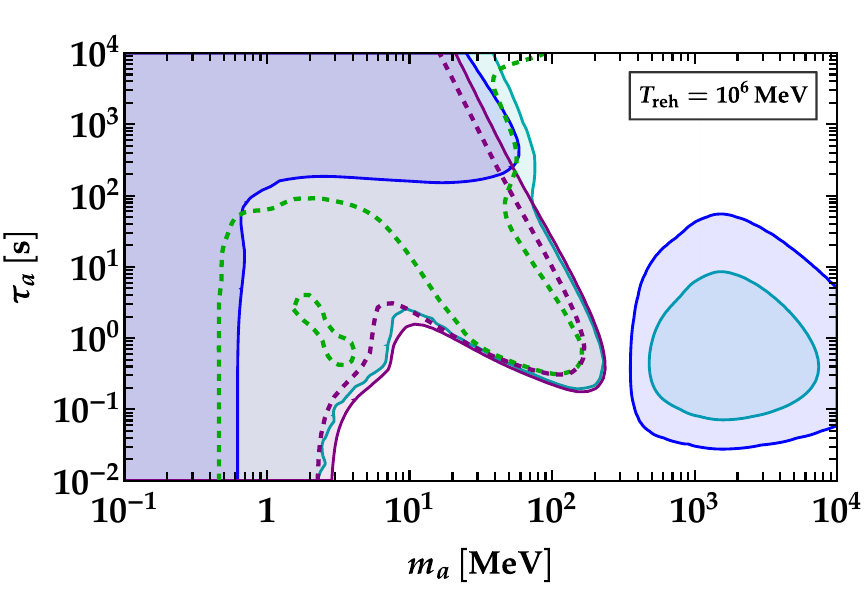}~\includegraphics[width=0.5\linewidth]{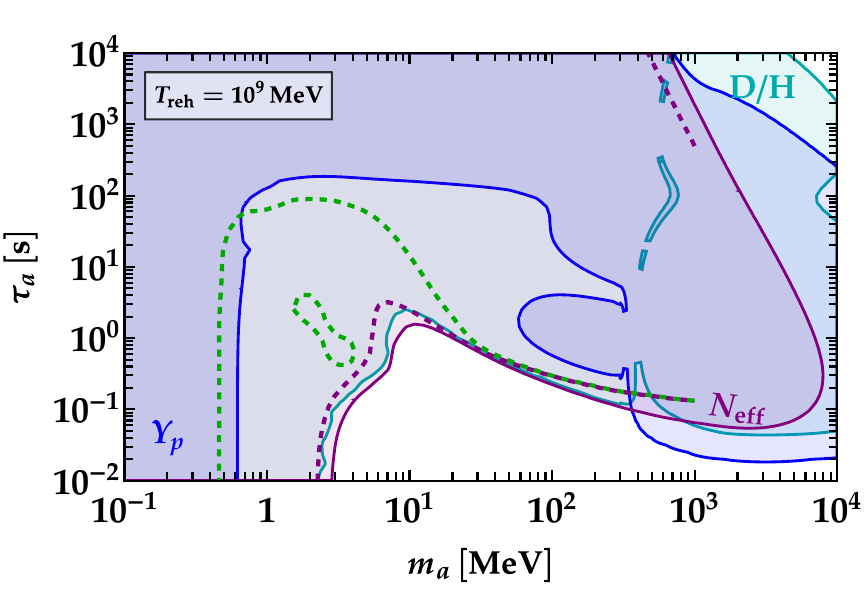}
    \caption{Comparison of the results of our work with Fig. 5 of Ref.~\cite{Depta:2020wmr}. As in Fig.~\ref{fig:comparison-cadamuro}, the shaded domains are our results, whereas the dashed lines denote the results of Ref.~\cite{Depta:2020wmr}, where the green domain denotes their combined \yp+\dH bounds and the purple the $N_{\rm eff}$ one. We compare the results for several choices of reheating temperature $\Treh = 10, 10^{3}, 10^{6}, 10^{9}$ MeV. }
    \label{fig:comparison-robust}
\end{figure}

Now, let us compare with Ref.~\cite{Depta:2020wmr}. This work in particular studies how the impact of the ALPs on cosmology changes with the reheating temperature \Treh. We will utilize their Fig.~5, where the values $\Treh = 10, 10^{3}, 10^{6}$, and $10^{9}\mev$ are considered. The comparison is shown in Fig.~\ref{fig:comparison-robust}.

By comparing the solid and dashed purple lines, we can clearly conclude that our evaluations of $N_{\rm eff}$ are quite similar. We attribute the small mismatch on the limit to the fact that (i) we are solving differently for neutrino decoupling, (ii) the number we use for the $N_{\rm eff}$ limit is slightly different, and (iii) our treatment of the ALP production in the early Universe is also different.

However, we find significantly different shapes for the joint BBN limit. First, considering the right top panel, we see that the limit obtained in Ref.~\cite{Depta:2020wmr} is significantly weaker than ours. In particular, the BBN exclusion shape actually quite closely follows the pure $Y_{\rm P}$ bounds as obtained in our calculations, whereas our BBN bounds are dominated by the primordial deuterium. This may suggest that the ${\rm D/H}|_{\rm P}$ limits were not included in the domain from Ref.~\cite{Depta:2020wmr}.

Second, our results from the left-lower panel contain the island at $m_a>2m_{\pi^\pm}$, which appears in our analysis as we are taking into account the effect from rare meson decays and their interactions. Similarly, in the lower-right panel, we see that our limits extend to smaller $\tau_a$ lifetimes. 

Finally, let us comment on the ALP lifetimes $\tau_{a}\gtrsim 10^{4}$, which are beyond the parameter space we considered in this work but have been covered in~\cite{Depta:2020wmr}. For such lifetimes, the ALPs may survive until keV temperatures, where their energetic EM decay products may dissociate primordial nuclei. To see this, consider the injected photons $\gamma$, having energy $E_{\gamma}\approx m_{a}/2$. The dominant thermalization process of such photons, preventing them from dissociating nuclei, is $\gamma+\gamma_{\text{bg}}\to e^{+}e^{-}$. It is instant for temperatures $T>m_{e}^{2}/(22E_{\gamma})$~\cite{Kawasaki:1994sc}. Solving this inequality for $E_{\gamma} = 2.22\mev$ (deuterium binding energy), we get $T\simeq 5\,{\rm keV}$ and corresponding cosmic time $t\gtrsim 5\cdot 10^{4}\s$ as the time when photodisintegration would become important. 

We do not incorporate the photodisintegration, yet it may be included within our framework. It is important that for the ALP mass range $m_{a}\gtrsim 2m_{\pi}$, the photodisintegration competes with the meson-driven processes until lifetimes $\tau \simeq 10^{5}-10^{6}\s$. The resulting nuclear abundances evolution may be non-trivial; in particular, while the meson-driven processes tend to increase \dH, the photodisintegration would decrease it. Therefore, to get an accurate prediction of the nuclear abundances evolution, one would need to simultaneously include these two processes.

\bibliography{main.bib}

\end{document}